\documentclass[useAMS,usenatbib]{mn2e}  
\usepackage {graphicx,subfig,aas_macros,float,amsmath}
\usepackage{ulem}
\usepackage[dvipsnames]{xcolor}  

\newcommand{\equ}[1]{eq.~(\ref{eq:#1})}
\newcommand{\equs}[1]{eqs.~(\ref{eq:#1})}
\newcommand{\equm}[1]{(\ref{eq:#1})}
\newcommand{\Equ}[1]{Eq.~(\ref{eq:#1})}

\newcommand{\equnp}[1]{eq.~\ref{eq:#1}}
\newcommand{\equsnp}[1]{eqs.~\ref{eq:#1}}
\newcommand{\equmnp}[1]{\ref{eq:#1}}
\newcommand{\se}[1]{\S\ref{sec:#1}}
\newcommand{\fig}[1]{Fig.~\ref{fig:#1}}
\newcommand{\figs}[1]{Figs.~\ref{fig:#1}}
\newcommand{\figss}[1]{\ref{fig:#1}}
\newcommand{\Fig}[1]{Figure~\ref{fig:#1}}
\newcommand{\Figs}[1]{Figures~\ref{fig:#1}}
\newcommand{\tab}[1]{Table~\ref{tab:#1}}
\newcommand{\be}{\begin{equation}}
\newcommand{\ee}{\end{equation}}
\newcommand{\bea}{\begin{eqnarray}}
\newcommand{\eea}{\end{eqnarray}}

\newcommand{\no}{\noindent}

\newcommand{\msun}{{\rm M}_\odot}

\newcommand{\ifm}[1]{\relax\ifmmode#1\else$\mathsurround=0pt #1$\fi}
\newcommand{\kms}{\ifmmode\,{\rm km}\,{\rm s}^{-1}\else km$\,$s$^{-1}$\fi}

\newcommand{\kpc}{\,{\rm kpc}}
\newcommand{\pc}{\,{\rm pc}}

\newcommand{\Myr}{\,{\rm Myr}}
\newcommand{\K}{\,{\rm K}}

\newcommand{\ltsima}{$\; \buildrel < \over \sim \;$}
\newcommand{\lsim}{\lower.5ex\hbox{\ltsima}}
\newcommand{\gtsima}{$\; \buildrel > \over \sim \;$}
\newcommand{\gsim}{\lower.5ex\hbox{\gtsima}}

\def\la{\langle}

\def\sy{\,M_\odot\, {\rm yr}^{-1}}

\def\cmc{\,{\rm cm}^{-3}}

\def\M*{M_{\rm *}}
\def\Mv{M_{\rm v}}
\def\Rv{R_{\rm v}}
\def\Vv{V_{\rm v}}
\def\tv{t_{\rm v}}

\def\Tv{T_{\rm v}}
\def\Tb{T_{\rm b}}
\def\Ts{T_{\rm s}}
\def\Rs{R_{\rm s}}
\def\rhob{\rho_{\rm b}}
\def\rhos{\rho_{\rm s}}

\def\cb{c_{\rm b}}
\def\cs{c_{\rm s}}

\def\Vs{V_{\rm s}}
\def\hb{h_{\rm b}}
\def\hs{h_{\rm s}}

\def\tsc{t_{\rm sc}}
\def\tcm{t_{\rm cool,\,mix}}
\def\Mb{M_{\rm b}}
\def\Ms{M_{\rm s}}

\def\Pi{\varpi_{_{\rm I}}}

\usepackage{color}

\begin{document} 

\large 

\title[KHI with Cooling in Cold Streams]
{Instability of Supersonic Cold Streams Feeding Galaxies IV: Survival of Radiatively Cooling Streams}

\author[Mandelker et al.] 
{\parbox[t]{\textwidth} 
{ 
Nir Mandelker$^{1,2}$\thanks{E-mail: nir.mandelker@yale.edu },
Daisuke Nagai$^{1,3}$,
Han Aung$^3$,
Avishai Dekel$^4$,
Yuval Birnboim$^4$,
Frank C. van den Bosch$^1$
} 
\\ \\ 
$^1$Department of Astronomy, Yale University, PO Box 208101, New Haven, CT, USA;\\
$^2$Heidelberger Institut f{\"u}r Theoretische Studien, Schloss-Wolfsbrunnenweg 35, 69118 Heidelberg, Germany;\\
$^3$Department of Physics, Yale University, New Haven, CT 06520, USA;\\
$^4$Centre for Astrophysics and Planetary Science, Racah Institute of Physics, The Hebrew University, Jerusalem 91904, Israel
}
\date{} 
 
\pagerange{\pageref{firstpage}--\pageref{lastpage}} \pubyear{0000} 
 
\maketitle 
 
\label{firstpage} 
 
\begin{abstract} 
We study the effects of Kelvin Helmholtz Instability (KHI) on the cold streams that feed massive halos at high redshift, generalizing our earlier results to include the effects of radiative cooling and heating from a UV background, using analytic models and high resolution idealized simulations. We currently do not consider self-shielding, thermal conduction or gravity. A key parameter in determining the fate of the streams is the ratio of the cooling time in the turbulent mixing layer which forms between the stream and the background following the onset of the instability, $\tcm$, to the time in which the mixing layer expands to the width of the stream in the non-radiative case, $t_{\rm shear}$. This can be converted into a critical stream radius, $R_{\rm s,crit}$, such that $\Rs/R_{\rm s,crit}=t_{\rm shear}/\tcm$. If $\Rs<R_{\rm s,crit}$, the non-linear evolution proceeds similarly to the non-radiative case studied by \citet{M19}. If $\Rs>R_{\rm s,crit}$, which we find to almost always be the case for astrophysical cold streams, the stream is not disrupted by KHI. Rather, background mass cools and condenses onto the stream, and can increase the mass of cold gas by a factor of $\sim 3$ within 10 stream sound crossing times. 
The mass entrainment induces thermal energy losses from the background and kinetic energy losses from the stream, which we model analytically. Roughly half of the dissipated energy is radiated away from gas with $T<5\times 10^4\K$, suggesting much of it will be emitted in Ly$\alpha$.
\end{abstract} 
 
 
\section{Introduction}
\label{sec:intro}

\smallskip
Understanding how galaxies acquire fresh gas to fuel ongoing star formation is one of the most important outstanding issues in the field of galaxy formation. This is particularly acute during the peak phase of galaxy formation, at $z\sim(1-4)$. During this epoch, massive galaxies of $\sim 10^{11}\msun$ in baryons are observed to have star-formation rates (SFRs) of $\sim 100 \sy$, with short gas depletion times of $t_{\rm dep}\sim 100\Myr$ \citep{Genzel06,Forster06,Elmegreen07,Genzel08,Stark08}. According to the standard $\Lambda {\rm CDM}$ cosmological model, such galaxies are predicted to reside in dark matter halos with virial mass $\Mv\gsim 10^{12}\msun$ \citep{Dekel09,Wechsler18,Behroozi19}. Such halos are predicted to support a stable accretion shock near the virial radius, $\Rv$, such that the gas within their circumgalactic medium (CGM) has temperatures of order the virial temperature, $\Tv\sim 10^6\K$, and cooling times of order the Hubble time \citep{Rees77,White78,bd03}. 
While the cooling time becomes shorter near the halo centre, it is unclear whether this can result in cooling flows capable of sustaining the high SFRs, which are comparable to the predicted cosmological accretion rate of gas onto the halos \citep{Dekel09}.
Since these star-forming galaxies appear to be extended rotating disks, it is also unlikely that mergers are responsible for driving the cold gas towards the galaxies \citep{Genzel06,Shapiro08,Forster09,wisnioski15,Simons19}.

\smallskip
In the contemporary picture of galaxy formation, such massive galaxies at high redshift are fed by narrow streams of dense gas which trace cosmic web filaments \citep{db06,Dekel09}. Owing to their high densities and short cooling times, the gas in these streams does not shock, maintains a temperature of $\Ts\gsim 10^4\K$, 
and is thought to be able to penetrate through the hot CGM and reach the central galaxy in roughly a virial crossing time. This is commonly referred to as the cold-stream model of galaxy formation. Such cold streams are ubiquitous in cosmological simulations of galaxy formation \citep[e.g.][]{Keres05,Ocvirk08,Dekel09,CDB,FG11,vdv11,Harford11}, where they are found to supply the halo with gas at rates comparable both to the predicted cosmological accretion rate and the observed SFR, implying that a significant fraction of the gas must reach the central galaxy \citep{Dekel09,Dekel13}. Their narrow size 
makes the streams difficult to directly detect observationally. However, numerous studies of the CGM around high-redshift massive galaxies, in both absorption \citep{Fumagalli11,Goerdt12,vdv12,Bouche13,Bouche16,Prochaska14} and emission \citep{Steidel00,Matsuda06,Matsuda11,Cantalupo14,Martin14a,Martin14b,Martin19,Borisova16,Fumagalli17,Leclercq17,Arrigoni18} reveal large quantities of cold gas with spatial and kinematic properties consistent with predictions for cold streams.


\smallskip
Despite the success of the cold-stream model, the evolution of streams in the CGM is still widely disputed. Different cosmological simulations make different predictions regarding whether or not the streams remain coherent and penetrate all the way to the central galaxy, how much of their energy is dissipated within the halo and how much radiation this dissipation produces, and what effects these may have on the thermal and morphological properties of the gas that eventually joins the galaxy. The latter is particularly important for the growth of galactic disks, as the streams are thought to play a key role in the buildup of angular momentum in disk galaxies \citep{Pichon11,Kimm11,Stewart11,Stewart13,Codis12,Danovich12,Danovich15}. 

\smallskip
Some of the differences between various simulations have been attributed to different numerical methods. For instance, simulations using the moving mesh code \texttt{AREPO} \citep{Springel10,Vogelsberger12} suggest that streams heat-up and dissolve at $\gsim 0.5\Rv$ \citep{Nelson13}, while comparable Eulerian AMR \citep{CDB,Danovich15} and Lagrangian SPH \citep{Keres05,FG10} simulations find that the streams remain cold and collimated until $\sim 0.25\Rv$. However, since none of the aforementioned simulations are able to properly resolve the streams in the outer halo, the interpretation of these results is uncertain. The resolution in most state-of-the-art simulations is adaptive in a quasi-Lagrangian sense, such that the effective mass resolution is fixed. Consequently, the spatial resolution becomes very poor in the low density CGM near the virial radius, with typical cell sizes of several hundred pc to kpc scales \citep[e.g.][]{Nelson16}. This is comparable to the stream width itself, which is expected to be of order a few kpc \citep{P18,M18}. Therefore, current cosmological simulations cannot resolve instabilities at smaller scales which are critical to properly model the detailed evolution of cold streams. While several groups have recently introduced new methods to better resolve the CGM \citep{Hummels18,Corlies18,Peeples19,Suresh19,vdv19}, these have focused on less massive galaxies, and have not yet been used to study cold streams. 

\smallskip
As an alternative to cosmological simulations, several recent studies have turned to analytic models and idealized, high-resolution numerical simulations to study the evolution of cold streams. In a series of papers, \citet{M16,P18} and \citet{M19} (hereafter M16; P18; and M19 respectively) studied the Kelvin-Helmholtz Instability (KHI) of a dense stream moving supersonically through a tenuous background in the linear (M16) and non-linear (P18; M19) regimes. These studies were purely hydrodynamical in nature, neglecting additional physical processes such as gravity, radiative cooling and heating, or magnetic fields. They found that sufficiently narrow streams should completely dissintegrate in the CGM prior to reaching the central galaxy. The smallest radius which would allow a stream to survive for a virial crossing time ranged from $(0.005-0.05)\Rv$, with denser and faster streams having smaller critical radii for disruption (M19). They also found that KHI caused streams to decelerate, and estimated that for typical stream parameters between $(10-50)\%$ of the gravitational potential energy gained by inflow down the dark-matter halo potential from $\Rv$ to $0.1\Rv$ can be dissipated in the CGM (M19). This may explain why some cosmological simulations find that streams maintain a roughly constant inflow velocity within the halo, rather than accelerate due to gravity \citep{Dekel09,Goerdt15a}. 

\smallskip
Following up on these studies, several other works have used similar idealized methods to investigate the impacts of additional physics such as self-gravity \citep{Aung19} or magnetic fields \citep{Berlok19b} on the evolution of cold streams. Others have addressed more specific questions, such as the impact of KHI on the density distribution within streams \citep{Vossberg19}, or the impact of a smooth transition between the stream and the background on the linear growth rate of KHI \citep{Berlok19a}, while still only considering adiabatic hydrodynamics. We comment on the results of these studies in \se{phys} below.

\smallskip
While the aforementioned studies have been extremely comprehensive, none have included radiative cooling or heating, which is the focus of the present study. Cooling can have a strong influence on KHI in both the linear and non-linear regimes \citep{Massaglia92,Massaglia96,Bodo93,Vietri97,Rossi97,Hardee97,Stone97,Xu00,Micono00}, and can either enhance or inhibit the growth rates depending on the cooling function. Additionally, studies of turbulent mixing layers, such as those formed as the result of KHI, show that cooling can qualitatively alter their evolution (\citealp{Gronke18,Gronke20,Ji19}, hereafter G18; G20; and J19 respectively; \citealp{Li19}). This has been shown to significantly extend the lifetime of cool clouds traveling through a hot background due to condensation of hot gas onto the cloud tail. This process may also be important for the evolution of cold streams (see \se{theory}). Finally, explicit treatment of radiative processes is necessary in order to address one of the key open questions regarding cold streams, namely whether the dissipation of their energy in the CGM is capable of powering the Ly$\alpha$ emission of observed Ly$\alpha$ blobs. These have typical luminosities of $\sim 10^{43}~{\rm erg~s^{-1}}$ and sizes of $\sim(50-100)\kpc$ \citep{Steidel00,Matsuda06,Matsuda11}. As noted above, M19 found that streams can dissipate up to $50\%$ of the gravitational energy gained by flowing down the halo potential due to KHI. If this energy is radiated away as Ly$\alpha$ then the resulting luminosities would be consistent with Ly$\alpha$ blobs \citep{Dijkstra09,Goerdt10,FG10}. However, since M19 did not include radiative cooling, the amount of energy which would be radiated away, as opposed to heating the stream or the background, remained unclear.

\smallskip
The remainder of this paper is organized as follows. In \se{theory} we asses when cooling may be important for the evolution of cold streams by comparing the results of M19 regarding stream disruption and deceleration due to non-radiative KHI with recent results on the evolution of radiative turbulent mixing layers. In \se{methods} we introduce a suite of numerical simulations used to study stream evolution. In \se{results} we present the results of our numerical analysis and compare these to our analytic predictions. 
In \se{phys} we discuss the potential effects of additional physics not included in our current analysis. 
We summarize our conclusions in \se{conc}.

\section{Theoretical Framework} 
\label{sec:theory} 

\smallskip
In this section, we begin in \se{theory_khi} by reviewing the main aspects of non-linear KHI in cylindrical streams without radiative cooling, in particular the formation of a turbulent mixing layer in between the two fluids. We continue in \se{theory_turb} by reviewing some general properties of turbulent mixing layers with and without radiative cooling. Finally, in \se{theory_cool}, we use the previous two subsections to predict when cooling will be important in the non-linear evolution of KHI in cold streams.

\subsection{Non-Radiative KHI in Cylindrical Streams}
\label{sec:theory_khi}

\smallskip
We focus here on the results of M19, who studied the non-linear evolution of KHI in a dense cylinder moving through a static and dilute background, expanding upon earlier work by M16 and P18. This system is characterized by two dimensionless parameters, the Mach number of the flow with respect to the sound speed in the background, $\Mb=V_{\rm s}/\cb$, and the density contrast between the stream and the background, $\delta=\rhos/\rhob$. If the stream and the background are in pressure equilibrium, the Mach number with respect to the sound speed in the stream is $\Ms=V_{\rm s}/\cs=\delta^{1/2}\Mb$. The stream radius is $\Rs$.

\smallskip
We begin by noting that there are two qualitatively different modes of KHI. \textit{Surface modes}, as their name suggests, are initially concentrated near the interface between the two fluids, and propagate into both fluids. In the early non-linear stages of evolution, they display the familiar cat's eye vortex morphology commonly associated with textbook KHI \citep[e.g.][]{Chandrasekhar61}. \textit{Body modes}, which are relevant at high Mach number flows where $V_{\rm s}>(\cs+\cb)$, are caused by sound waves propagating inside the stream and being reflected off its boundaries while undergoing constructive interference. This causes a global, large scale, deformation of the stream, while leaving the interface itself relatively unperturbed until the final stages of stream disruption. In two dimensions, where the stream is not a cylinder but rather a planar slab confined to the region $-\Rs<x<\Rs$ while flowing along the $z$ direction, surface modes stabilize once $\Mb$ is larger than a critical value of order unity, leaving only unstable body modes. Therefore, the distinction between these two modes is critical (M16; P18). However, in three dimensional cylinders, surface modes with sufficiently large azimuthal wave-numbers, $m=2\pi\Rs/\lambda_{\varphi}$, are always unstable.\footnote{The same is true for planar slabs in three dimensions, where surface modes with sufficiently large wavenumbers perpendicular to the flow yet within the slab plane are always unstable.} Body modes, which evolve on longer timescales than surface modes, thus rarely manifest in practice (M19). We therefore restrict the following discussion to surface modes.

\smallskip
The non-linear evolution of KHI surface modes is characterized by the formation of a turbulent shear layer in between the two fluids. This layer forms at the initial interface between the fluids and expands into both fluids, growing by the merger of eddies which drives power towards larger and larger scales. At the same time, the largest eddies break up and transfer power towards smaller scales, generating turbulence. The width of the shear layer as a function of time can be well approximated by 
\be 
\label{eq:non_rad_h_t}
h(t) = \alpha V_{\rm s} t,
\ee
{\no}where the dimensionless growth rate, $\alpha$, 
can be approximated as \citep{Dimotakis91} 
\be 
\label{eq:alpha_non_rad}
\alpha \simeq 0.21\times \left[0.8{\rm exp}\left(-3 M_{\rm tot}^2\right)+0.2\right],
\ee
with $M_{\rm tot}=\Vs/(\cs+\cb)$. This approximation is an excellent fit for 2d planar slabs and for 3d cylinders at early times. At late times, once $h\sim \Rs$, \equ{non_rad_h_t} can still describe shear layer growth in 3d cylinders with a modified value of $\alpha$, which differs from \equ{alpha_non_rad} by a factor of $\lsim 2$ (M19). 

\smallskip
The shear layer expands asymmetrically into the stream and the background, due to their different densities. The stream penetration depth into either fluid can be shown to be 
\be 
\label{eq:hs_growth}
\hs(t) = \frac{1}{1+\sqrt{\delta}}h(t),\quad \hb(t) = \frac{\sqrt{\delta}}{1+\sqrt{\delta}}h(t),
\ee
{\no}such that $h=h_{\rm s}+h_{\rm b}$. Stream disruption was said to occur when $\hs=\Rs$ at time
\be 
\label{eq:tdis_non_rad}
t_{\rm dis}=\frac{\left(1+\sqrt{\delta}\right)\Rs}{\alpha\Vs}.
\ee
{\no}However, in \citet{Aung19} it was found that when comparing the timescale for KHI induced stream disruption to the timescale for gravitational instability induced stream fragmentation, the relevant KHI timescale was that for which $h=\Rs$, namely when the total width of the shear layer exceeds the stream radius. This is sensible, because KHI is driven by the presence of a contact discontinuity between the fluids, which disappears once $h\simeq \Rs$. This is also the time when shear layer growth in 3d cylinders begins to deviate from \equ{alpha_non_rad} (M19). This occurs at time 
\be 
\label{eq:tshear_non_rad}
t_{\rm shear} = \frac{\Rs}{\alpha \Vs}.
\ee 
{\no}An additional important timescale is the stream sound crossing time,
\be 
\label{eq:tsc}
\tsc = \frac{2\Rs}{\cs}.
\ee 

\smallskip
As the shear layer expands into the background, it entrains more and more background mass which shares the initial stream momentum. This causes the stream fluid in the shear layer to decelerate, and its velocity as a function of time is well fit by (M19) 
\be 
\label{eq:stream_deceleration}
V_{\rm s}(t) = \frac{V_{\rm s,0}}{1+t/t_{\rm dec}},
\ee
{\no}with $V_{\rm s,0}$ the initial velocity of the stream, and
\be 
\label{eq:tau_dec}
t_{\rm dec} = \dfrac{\left(1+\sqrt{\delta}\right)\left(\sqrt{1+\delta}-1\right)}{\alpha\sqrt{\delta}}\frac{\Rs}{V_{\rm s,0}},
\ee
{\no}the time when the entrained background mass equals the initial stream mass. 

\smallskip
Since the deceleration is induced by mass entrainment and momentum conservation, we have that $mV=m_0 V_0={\rm const}$, with $m$ the total mass per unit length, stream plus background, that is flowing and $V$ is the characteristic velocity of the flow. Since the total kinetic energy per unit length associated with this laminar flow is $E_{\rm k}\sim 0.5m V^2\sim E_{\rm k,0}V/V_0$, the bulk kinetic energy decreases with time as 
\be 
\label{eq:Ek_diss_non_rad}
E_{\rm k}(t) = \frac{E_{\rm k,0}}{1+t/t_{\rm dec}},
\ee 
{\no}similar to the stream velocity. This dissipated energy can be converted into turbulence or into thermal energy in either the stream or the background, or be advected away from the stream via sound waves. 
If the gas can cool, some fraction of it will be radiated away.

\subsection{Turbulent Mixing Layers}
\label{sec:theory_turb}

\smallskip
Radiatively cooling turbulent mixing layers have received much interest in the literature lately, primarily in the context of cold clouds moving through a hot medium either in the ISM or CGM (G18; G20; J19). It has been found that cooling in these layers can have a profound impact on the survival of such clouds. Since KHI surface modes in cylindrical streams also form a turbulent mixing layer, it is likely that radiative cooling can have a similar influence on stream survival. We here summarize some of the general properties of radiative turbulent mixing layers, modifying them for cylindrical rather than initially spherical or planar geometry where relevant (and stating so explicitly). We apply these results to cold streams in the next subsection.

\smallskip
The general structure of turbulent mixing layers was studied analytically by \citet{Begelman90} and recently revisited by G18. At the interface between hot and cold gas in pressure equilibrium, with densities and temperatures $\rho_{\rm b,s}$ and $T_{\rm b,s}$, respectively, KHI will cause a turbulent mixing layer to develop. By estimating the mass entrainment rates of hot and cold gas into the mixing layer, they derive the mean density and temperature of the mixing layer to be 
\be 
\label{eq:nmix}
\rho_{\rm mix}\simeq \left(\rhob \rhos\right)^{1/2} = \delta^{-1/2}\rhos,
\ee 
\be 
\label{eq:Tmix}
T_{\rm mix}\simeq \left(\Tb \Ts \right)^{1/2} = \delta^{1/2}\Ts.
\ee

\smallskip
An important timescale in determining the dynamics of such systems is the radiative cooling time of the mixing layer, 
\be 
\label{eq:tcool_mix}
\tcm = \frac{k_{\rm B}T_{\rm mix}}{(\gamma-1)n_{\rm mix}\Lambda(T_{\rm mix})},
\ee
{\no}where $\gamma$ is the adiabatic index of the gas, $k_{\rm B}$ is Boltzmann's constant, $n_{\rm mix}$ is the particle number density in the mixing region, and $\Lambda(T_{\rm mix})$ is the cooling function evaluated at $T_{\rm mix}$.

\smallskip
In their study of the evolution of cold clouds moving supersonically in a hot atmosphere, G18 and G20 find qualitatively different results depending on whether $\tcm$ is larger or smaller than the cloud-crushing time, $t_{\rm cc}\simeq \delta^{1/2}r_{\rm cl}/V_{\rm cl}$, where $r_{\rm cl}$ is the cloud radius and $V_{\rm cl}$ is its velocity (though see \citealp{Li_hopkins19} for a different formalism involving the cooling time in the hot medium). 
$t_{\rm cc}$ is the characteristic timescale for the cloud to be destroyed by a combination of KHI, Rayleigh-Taylor instabilities, and ram-pressure shocks crossing the cloud. If $\tcm>t_{\rm cc}$ then while the cloud lifetime is extended by a factor of a few compared to the non-radiative case, it is ultimately still disrupted and mixed into the background within a few $t_{\rm cc}$ (\citealp{Scannapieco15,Schneider17}; G18). However, if $\tcm<t_{\rm cc}$, then the cloud not only survives but actually grows in mass by entraining background material (\citealp{Armillotta16,Armillotta17}; G18; G20; \citealp{Li_hopkins19}). In the former case, KHI transports cold gas from the cloud through the turbulent mixing layer and mixes it into the background. In the latter case, the flow through the turbulent mixing layer is driven by pressure gradients formed as a result of radiative cooling within the layer, and the net result is background gas cooling and condensing onto the cold cloud (J19; G20). 

\smallskip
The accretion rate of background gas onto the cold cloud can be written as 
\be 
\label{eq:mdot}
\dot{m}\simeq \rhob A v_{\rm mix},
\ee
{\no}where $A$ is the effective surface area of the mixing layer and $v_{\rm mix}$ is the characteristic velocity of material flowing through the mixing layer. In the case of initially spherical clouds, the surface area grows much larger than $4\pi r_{\rm cl}^2$ due to the formation of an extended tail on the trailing end. It can be shown that in this case $A\propto r_{\rm cl}^2\delta$ (G20). However, this will not be the case for an initially cylindrical stream since there is no tail formation. In this case, the surface area will scale as  
\be 
\label{eq:area}
A\sim 2\pi\Rs L
\ee 
{\no}where $L$ is the length of the stream. For an infinite stream, $m$ and $A$ can be taken per unit length. 
Note that in practice, the actual surface area of the mixing layer can be much larger, due to small scale fractal structures that develop in the turbulent mixing zone. Such small scale structures are difficult to account for analytically, and in the cloud simulations of G20, their total surface area remains unconverged with 64 cells across the cloud radius. Numerically evaluating the total surface area in our simulations is beyond the scope of this paper. Rather, similarly to G20, we use \equ{area} as an effective surface area for the mixing layer on the large scales which drive the mixing. We comment more on this below.

\smallskip
The mixing velocity, $v_{\rm mix}$, scales as (G20) 
\be 
\label{eq:vmix_gronke}
v_{\rm mix}\propto \cs (t_{\rm cool,s}/t_{\rm sc})^{-1/4},
\ee
{\no}where $t_{\rm cool,s}$ is the cooling time in the cold phase. The scaling $v_{\rm mix}\propto t_{\rm cool}^{-1/4}$ was first seen in the turbulent mixing layer simulations of J19, and was later explained by G20. For completeness, we repeat the main arguments of the derivation here, referring readers to J19 and G20 for a more in depth discussion. The basic idea rests on two assumptions. First, that the growth of the mixing layer can be modeled as a turbulent diffusion process with the size of the mixing layer, $H$, determined by equating the diffusion and cooling timescales. This yields $H\propto (r_{\rm cl}\cs t_{\rm cool})^{1/2}$. Second, that within the mixing layer, pressure fluctuations are generated by cooling at a rate $\delta {\dot{P}}\sim P/t_{\rm cool}$, damped by sound waves at a rate $\delta {\dot{P}}\sim -\delta P/t_{\rm sc,H}$, and that these balance each other in steady-state. This yields $\delta P/P\sim t_{\rm sc,H}/t_{\rm cool}\sim H/(\cs t_{\rm cool})\propto [r_{\rm cl}/(\cs t_{\rm cool})]^{1/2} \sim (\tsc/t_{\rm cool})^{1/2}$. In this case, the inflow into the mixing layer from large scales is dominated by a quasi-isobaric cooling flow obeying a Bernoulli-like constraint $P+\rho v_{\rm mix}^2\sim {\rm const}$, from where it follows that $v_{\rm mix}^2\sim \delta P/\rho$ and hence $v_{\rm mix}\sim \cs (\tsc/t_{\rm cool})^{1/4}$.

\smallskip
As we will discuss in \se{theory_cool}, we assume that the stream is in thermal equilibrium with a UV background, so the net cooling time within the stream is infinite. As described above, the cooling time enters the expression for $v_{\rm mix}$ through the formation of pressure gradients, which will be largest where the cooling time is minimal. The minimal net cooling time turns out to always occur at roughly $T\sim 1.5\Ts$, so we hereafter replace $t_{\rm cool,s}$ in \equ{vmix_gronke} with $t_{\rm cool,T=1.5\Ts}$. This assumption will be validated using numerical simulations in \se{results}.

\smallskip
In the cloud simulations of G20, the amplitude of $v_{\rm mix}$ is found to be unconverged with 64 cells across the cloud radius, similar to $A$, due to the fractal structure of the mixing layer on small scales. However, \equ{vmix_gronke} is found to be representative of the velocity through the mixing layer on large scales, just as \equ{area} is representative of the surface area on large scales.

\smallskip
Combining \equs{mdot}-\equm{vmix_gronke} we obtain an expression for the mass accretion rate onto a cylindrical stream 
\be 
\label{eq:mdot_gronke}
\dot{m} \sim \frac{4}{\delta}\frac{m_{\rm 0}}{t_{\rm sc}}\left(\frac{t_{\rm cool,1.5\Ts}}{t_{\rm sc}}\right)^{-1/4}={\rm const},
\ee
{\no}where $m_{\rm 0}=\pi\Rs^2 L \rhos$ is the initial stream mass, and $\tsc$ is given by \equ{tsc}. Note the additional factor $\delta^{-1}$ compared to the corresponding expression in G20, due to the different scaling of the surface area, $A$, discussed above. 

\smallskip
Unlike the surface area of and velocity through the mixing layer, $A$ and $v_{\rm mix}$, the mass flux through the mixing layer is found to be converged with as few as 8 cells per cloud radius in the simulations of G20. We find the same for our simulations of cylindrical streams (Appendix \se{convergence}). We interpret this as the result of a turbulent cascade within the mixing layer, whereby the flux of mass and energy towards smaller scales is independent of scale, as in Kolmogorov turbulence. The total mass flux through the mixing layer is thus set by the large scales which drive the mixing, and not the small scales which are dominated by the fractal structure. This justifies our use of \equs{area} and \equm{vmix_gronke} in the expression for ${\dot{m}}$ in \equ{mdot_gronke}. In \se{results} we will validate these assumptions by comparing \equ{mdot_gronke} to simulation results. However, a scale-dependent measurement of $A$ and $v_{\rm mix}$ in our simulations is beyond the scope of this paper, and left for future work which will characterise the nature of the turbulence in the stream-background interface.

\begin{figure*}
\begin{center}
\includegraphics[trim={0.5cm 0.0cm 0.8cm 0.0cm}, clip, width =0.98 \textwidth]{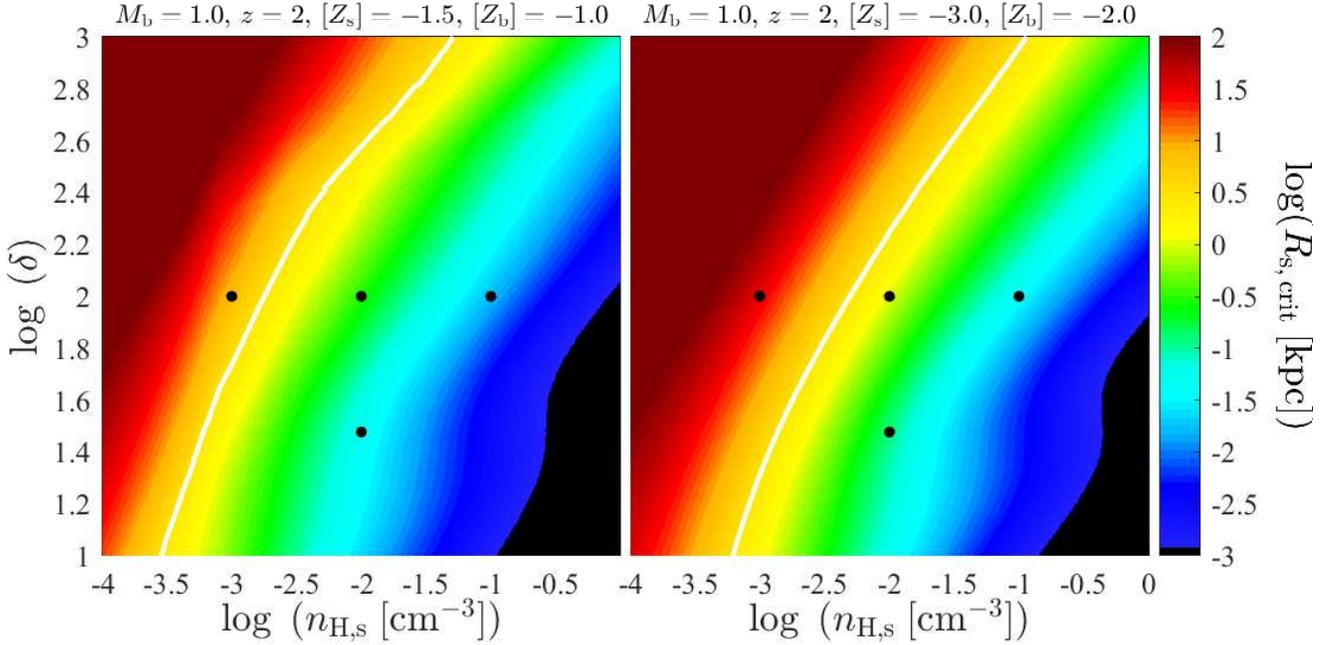}
\end{center}
\caption{Critical stream radius for which $t_{\rm cool,mix}=t_{\rm shear}$, as a function of stream density, $n_{\rm H,s}$, and density contrast with respect to the halo, $\delta$, following \equ{Rscrit}. The stream temperature is evaluated by assuming thermal equilibrium between the stream and a \citet{HaardtMadau96} UV background at $z=2$. We further assume $\Mb=1$ and evaluate $\alpha$ from \equ{alpha_non_rad}. The left (right) panels represent upper and lower limits for the metalicity in the stream and background, with assumed values listed in each panel. The white contours mark $R_{\rm s,crit}=3\kpc$, the fiducial value of $\Rs$ used in our simulations. Black points mark $(n_{\rm H,s},\delta)$ values used in our simulations.}
\label{fig:Rs_crit} 
\end{figure*}

\subsection{The Importance of Cooling for Cold Stream Stability}
\label{sec:theory_cool}

\smallskip
In the case of a cylindrical stream in a shear flow, the time when the shear layer expands to the size of the stream, $t_{\rm shear}$ (\equnp{tshear_non_rad}), can be considered analogous to the cloud crushing time. Therefore, motivated by the results discussed above, we speculate that when $\tcm\gsim t_{\rm shear}$, the non-linear evolution of KHI in streams will be similar to the non-radiative case, where streams disrupt and mix into the background. However, if $\tcm\lsim t_{\rm shear}$, cold streams may survive for much longer timescales and even grow in mass. Equating $\tcm=t_{\rm shear}$ gives a critical stream radius, 
\be 
\label{eq:Rscrit}
R_{\rm s,crit}\simeq 0.3\kpc~ \alpha_{0.1}\,\delta_{100}^{3/2}\,\Mb\,\frac{T_{\rm s,4}}{n_{\rm s,0.01}\Lambda_{\rm mix,-22.5}},
\ee
{\no}where $T_{\rm s,4}=\Ts/10^4\K$, $n_{\rm s,0.01}=n_{\rm s}/0.01\cmc$, $\Lambda_{\rm mix,-22.5}=\Lambda(T_{\rm mix})/10^{-22.5}{\rm erg~s^{-1}~cm^3}$, $\delta_{100}=\delta/100$, and $\alpha_{0.1}=\alpha/0.1$. Since $\tcm/t_{\rm shear}=(\Rs/R_{\rm s,crit})^{-1}$, streams with $\Rs\gsim R_{\rm s,crit}$ are expected to grow in mass rather than dissolve. We note, however, that the critical ratio of $\tcm/t_{\rm shear}$ need not be precisely unity, yielding some uncertainty in the precise normalization of $R_{\rm s,crit}$. We will examine this further below using numerical simulations.

\smallskip
In \fig{Rs_crit}, we present $R_{\rm s,crit}$ as a function of the Hydrogen number density in the stream\footnote{For fully ionized gas with primordial composition, i.e a Hydrogen mass fraction of $X=0.76$ and a mean molecular weight of $\mu\sim 0.59$, the total number density used in \equ{Rscrit} is related to the Hydrogen number density by $n_{\rm s}\simeq 2.2n_{\rm H,s}$.} $n_{\rm H,s}$, and the density contrast, $\delta$. To evaluate the stream temperature, we assume the stream to be in thermal equilibrium with a $z=2$ \citet{HaardtMadau96} UV background. In particular, we are ignoring the possibility that the streams may be self-shielded. This equilibrium temperature varies with density, but is roughly $T_{\rm s,4}\sim (1-2)$. We assume $\Mb=1$ and use \equ{alpha_non_rad} to evaluate $\alpha$. The cooling curve depends on the metalicity in the stream and the background. These are not well constrained, but simulations suggest that $Z_{\rm s}$ can range from $\sim 10^{-3}$ to a few percent of the solar value, while $Z_{\rm b}$ ranges from $\sim (0.01-0.1)Z_{\odot}$ \citep[e.g.][]{Goerdt10,Fumagalli11,vdv12b,Roca19}. To bracket these two extremes, we show on the left-hand side of \fig{Rs_crit} a model with $Z_{\rm s}=0.03Z_{\odot}$ and $Z_{\rm b}=0.1Z_{\odot}$, while on the right-hand side we assume $Z_{\rm s}=0.001Z_{\odot}$ and $Z_{\rm b}=0.01Z_{\odot}$. In each case we assume the metalicity in the mixing region is $Z_{\rm mix}=(Z_{\rm s}Z_{\rm b})^{1/2}$, in accordance with \equ{nmix}. 

\smallskip
Analytic considerations and cosmological simulations suggest that for cold streams feeding $\sim 10^{12}\msun$ halos at redshifts $z=(2-3)$, $n_{\rm H,s}\sim (0.001-0.1)$ \citep{Goerdt10}, $\delta\sim (30-300)$\footnote{Note that this is slightly higher than the typical range of $\delta=(10-100)$ assumed in M16, P18, and M19}
and $\Mb\sim (0.5-2.0)$ (P18; M19). In this range, we have $R_{\rm s,crit}\sim (0.01-100)\kpc$, spanning four orders of magnitude, increasing as either the density or metalicity in the stream or the background decrease. While the actual stream radii are not well constrained, analytic considerations predict (P18)
\be 
\label{eq:Rs_Rv}
\Rs/\Rv\sim (0.01-0.1)~\left(n_{s,0.01}\,\Mb\right)^{-1/2},
\ee 
{\no}with a typical value of $\sim 0.06$. The right-hand-side of \equ{Rs_Rv} can be inferred from equations (62) and (68) of P18. The range $(0.01-0.10)$ comes from uncertainties in model parameters (see P18). The virial radius is \citep[e.g.][]{Dekel13}
\be 
\label{eq:Rvir}
\Rv\simeq 100\kpc~M_{12}^{1/3}\,(1+z)_3^{-1},
\ee
{\no}with $M_{12}=\Mv/10^{12}\msun$ and $(1+z)_3=(1+z)/3$. This implies that in halos with $\Mv\gsim 10^{12}\msun$ at $z\sim (2-3)$, we expect stream radii in the range $\Rs\sim (0.5-10)\kpc\,n_{\rm s,0.01}^{-1/2}$. This can be smaller or larger than $R_{\rm s,crit}$ depending on the stream parameters, though we almost always expect $\Rs>R_{\rm s,crit}$. 

\smallskip
Note that $\Rs$ in the above discussion refers to the stream radius upon entering the halo at $r=\Rv$. As the stream penetrates into the halo, its density increases and its radius decreases. For an isothermal halo, we have $n\propto r^{-2}$ and $\Rs\propto r\propto n^{-1/2}$ while $\delta$ remains constant (P18;M19). Combined with \equ{Rs_Rv}, we have that $\Rs\propto n^{-1}$, similar to $R_{\rm s,crit}$. Thus, if $\Rs>R_{\rm s,crit}$ near $\Rv$, it is likely to remain so throughout the halo. In a companion paper \citep{M20}, we present a more detailed analytic model for stream evolution within dark matter halos.

\smallskip
The white contour in each panel of \fig{Rs_crit} marks $R_{\rm s,crit}=3\kpc$, which is the fiducial stream radius in the simulations described in \se{methods} below. In the region of $(n_{\rm H,s},\delta)$ space to the right of the curve, which represents most of the expected parameter range for cold streams, $\tcm<t_{\rm shear}$, and the non-linear behaviour of the turbulent mixing zone between the stream and the background will be dominated by cooling. To the left of the curve, we expect the non-linear behaviour to be qualitatively similar to the non-radiative case described in \se{theory_khi}. Black points mark $(n_{\rm H,s},\delta)$ values used in the simulations described in the next section.

\section{Numerical Methods} 
\label{sec:methods} 

\smallskip
In this section we describe the details of our simulation code and setup. We use the Eulerian AMR code \texttt{RAMSES} \citep{Teyssier02}, with a piecewise-linear reconstruction using the MonCen slope limiter \citep{vanLeer77}, and an HLLC approximate Riemann solver \citep{Toro94}. We utilise the standard \texttt{RAMSES} cooling module, which accounts for atomic and fine-structure cooling for our assumed metalicity values (see below), as well as photoheating and photoionization from a $z=2$ \citet{HaardtMadau96} UV background. We do not include self-shielding of dense gas in this work.

\subsection{Stream Parameters}
\label{sec:params}

\begin{table*}
\centering
\begin{tabular}{c|c|c|c|c|c|c|c|c|c}
${\rm Label}$ & $\Mb$ & $\delta$ & $n_{\rm H,s}$ & $\Rs$ & $\Ts$ & $T_{\rm max}$ & $m_{\rm max}$ & $t_{\rm cool,mix}$ & $t_{\rm cool,1.5\Ts}$ \\
 &  &  & $[\cmc]$ & $[\kpc]$  & $[10^4 \K]$ & $[10^4\K]$ &  & $[t_{\rm shear}]$ & $[t_{\rm sc}]$ \\
\hline
M0.5\_d100 & 0.5 & 100 & 0.01  & 3 & 1.51 & 20 & 1 & 0.114 & 0.0085 \\
M1.0\_d100 & 1.0 & 100 & 0.01  & 3 & 1.51 & 20 & 1 & 0.097 & 0.0085 \\
M2.0\_d100 & 2.0 & 100 & 0.01  & 3 & 1.51 & 20 & 1 & 0.145 & 0.0085 \\
M0.5\_d30  & 0.5 & 30  & 0.01  & 3 & 1.51 & 11 & 1 & 0.018 & 0.0085 \\
M1.0\_d30  & 1.0 & 30  & 0.01  & 3 & 1.51 & 11 & 1 & 0.016 & 0.0085 \\
M2.0\_d30  & 2.0 & 30  & 0.01  & 3 & 1.51 & 11 & 1 & 0.022 & 0.0085 \\
\hline
M1.0\_d100\_LD   & 1.0 & 100 & 0.001 & 3   & 1.76 & 20 & 1 & 2.441 & 0.6389 \\
M1.0\_d100\_HD   & 1.0 & 100 & 0.1   & 3   & 1.25 & 20 & 1 & 0.006 & 0.0002 \\
M1.0\_d100\_HR   & 1.0 & 100 & 0.01  & 6   & 1.51 & 80 & 1 & 0.049 & 0.0042 \\
M1.0\_d100\_LDHR & 1.0 & 100 & 0.001 & 30  & 1.76 & 80 & 1 & 0.244 & 0.0639 \\
M1.0\_d100\_LR   & 1.0 & 100 & 0.01  & 0.3 & 1.51 & 80 & 1 & 0.97 & 0.085 \\
\hline
M1.0\_d100\_HT & 1.0 & 100 & 0.01  & 3 & 1.51 & 80 & 1 & 0.097 & 0.0085 \\
M1.0\_d30\_HT  & 1.0 & 30  & 0.01  & 3 & 1.51 & 40 & 1 & 0.016 & 0.0085 \\
M2.0\_d100\_Hm & 2.0 & 100 & 0.01  & 3 & 1.51 & 80 & 4 & 0.145 & 0.0085 \\
\hline
M0.5\_d100\_NR & 0.5 & 100 & $--$ & $--$ & $--$ & $--$ & $--$ & $--$ \\
M1.0\_d100\_NR & 1.0 & 100 & $--$ & $--$ & $--$ & $--$ & $--$ & $--$ \\
M2.0\_d100\_NR & 2.0 & 100 & $--$ & $--$ & $--$ & $--$ & $--$ & $--$ \\
\end{tabular}
\caption{Stream parameters in the different simulations. The leftmost column is a label identifying the simulation, specifying the Mach number, the density contrast, and an additional comment. LD means low density (compared to the fiducial $n_{\rm H,s}$), HD means high density, HR means high (large) radius, LR means low (small) radius, HT means high $T_{\rm max}$, Hm means high (large) azimuthal mode numbers $m$ in the initial perturbations, and NR means non-radiative. Note that the HR, LDHR, LR, and Hm runs also have high $T_{\rm max}$. The next four columns list the control parameters: the Mach number, $\Mb$, density contrast, $\delta$, Hydrogen number density in the stream, $n_{\rm H,s}$ in $\cmc$, and the stream radius, $\Rs$ in $\kpc$. The next two columns list the stream temperature, $T_{\rm s}$ and the temperature above which cooling is shut off, $T_{\rm max}$, both in $\K$. 
The next column lists the maximal azimuthal mode number, $m$, with all modes $m=0-m_{\rm max}$ present in the initial perturbation spectrum. 
The final two columns list the ratios $\tcm/t_{\rm shear}$ and $t_{\rm cool,1.5\Ts}/\tsc$ (see \equ{vmix_gronke}). Only the LD simulation has $\tcm/t_{\rm shear}>1$, placing it in the slow-cooling regime (see \fig{Rs_crit}).}
\label{tab:sims}
\end{table*}

\smallskip
We perform a total of 17 simulations exploring 11 combinations of values for the stream Mach number, $\Mb=(0.5,1.0,2.0)$, density contrast, $\delta=(30,100)$, density, $n_{\rm H,s}=(0.001,0.01,0.1)\cmc$, and radius, $\Rs=(0.3,3,6,30)\kpc$. Three of these simulations were repeated with cooling switched off\footnote{Note that the non-radiative simulations are scale free, so the only meaningful parameters are $\Mb$ and $\delta$.}. The full list of simulations can be found in \tab{sims}. All simulations assume metalicity values $Z_{\rm s}=0.03Z_{\odot}$ and $Z_{\rm b}=0.1Z_{\odot}$. The equation of state of both fluids is that of an ideal monoatomic gas with adiabatic index $\gamma=5/3$. The initial stream temperature is set to be in thermal equilibrium with the UV background at the assumed stream density, and is 
typically $T_{\rm s}\sim 1.5\times 10^4\K$. The density in the background is $n_{\rm H,b}=\delta^{-1}n_{\rm H,s}$ and the temperature is determined by pressure equilibrium by $\Tb/\mu_{\rm b}=\delta\,\Ts/\mu_{\rm s}$, where $\mu_{\rm b}\sim \mu_{\rm s} \sim 0.6$ are the mean molecular weights in the background and stream respectively, determined from the ionization state.

\smallskip
To prevent the hot gas from cooling over long periods of time, we turn off cooling for gas with $T>T_{\rm max}$. In most of our simulations we use $T_{\rm max}\gsim(\Ts \Tb)^{1/2}$. We perform two additional simulations increasing $T_{\rm max}$ by a factor $\sim 4$ (see \tab{sims}) and find our results robust to this choice. This is consistent with the results of G18 and G20 which are shown to be robust to shutting off cooling at $T>0.6T_{\rm b}$ and not shutting it off at all. As we discuss in \se{results}, this is because most of the emissivity originates in cool gas with $T\gsim \Ts$, as also found in J19 and G20.

\smallskip
We run each of or simulations for 10 stream sound crossing times (\equnp{tsc}), saving 15 outputs per sound crossing time. The ratio of halo crossing time, $\tv=\Rv/V_{\rm s}$, to stream sound crossing time is 
\be 
\label{eq:tv_tsc}
\frac{\tv}{\tsc} = \frac{1}{2}\,\left(\frac{\Rs}{\Rv}\right)^{-1}\,\delta^{-1/2}\,\Mb^{-1}.
\ee 
{\no}For our simulated parameters, this is $\tv\sim (0.5-6)\tsc$.

\subsection{Simulation Domain, Boundary Conditions, \& Resolution}
\label{sec:bc}

\smallskip
The simulation domain is a cube of side $L=32\Rs$, extending from $-16\Rs$ to $16\Rs$ in all directions. We hereafter adopt the standard cylindrical coordinates, $(r,\varphi,z)$, where the $z$ axis corresponds to the stream axis. The stream fluid occupies the region $r<\Rs$ while the background fluid occupies the rest of the domain. 

\smallskip
We use periodic boundary conditions at $z=\pm 16\Rs$, and outflow boundary conditions, at $x=\pm 16\Rs$ and $y=\pm 16\Rs$, such that gas crossing the boundary is lost from the simulation domain. At these boundaries, the gradients of density, pressure, and velocity are set to 0. 

\smallskip
We use a statically refined grid with resolution decreasing away from the stream axis. The highest resolution region is ${\rm max}(|x|,|y|)<1.5\Rs$, with cell size $\Delta=\Rs/64$. The cell size increases by a factor of 2 every $1.5\Rs$ in the $x$ and $y$-directions, up to a maximal cell size of $\Rs/4$. The resolution is uniform along the $z$ direction, parallel to the stream axis. In the non-radiative case, KHI surface modes are converged at this resolution, both in terms of the cell size and the width of the high-resolution region (M19). Our main results for the radiative case are converged as well (appendix \se{convergence}).

\smallskip
Note that unlike the simulations presented in M16; P18; M19 and \citet{Aung19}, we do not smooth the interface between the stream and the background, maintaining instead a step function in $r$. Such a smooth transition layer is out of thermal equilibrium, and it rapidly cools and condenses onto the stream. Not only does this negate the initial purpose of smoothing, it also creates a weak shock wave that propagates towards the stream axis. While the lack of such a smooth transition makes the simulations susceptible to numerical perturbations on the grid scale, we have verified that these do not change the non-linear behaviour, since in any case we initialize a broad spectrum of perturbation modes (see also P18). 

\subsection{Perturbations} 
\label{sec:pert}

\smallskip
The stream is initialised with velocity ${\bf{V}}_{\rm s} = \Mb\cb {\bf{\hat{z}}}$, where $c_b=(\gamma P_{\rm b}/\rhob)^{1/2}$ is the sound speed in the background. The background gas is initialised at rest, with velocity ${\bf{V}}_{\rm b}=0$. 

\smallskip
We then perturb our simulations in a manner very similar to M19 and \citet{Aung19}. We initialize a random realization of periodic perturbations in the radial component of the velocity, $v_{\rm r}=v_{\rm x}{\rm cos}(\varphi)+v_{\rm y}{\rm sin}(\varphi)$. In practice, we perturb the Cartesian components of the velocity, 
\be 
\label{eq:pertx}
\begin{array}{c}
v_{\rm x}^{\rm pert}(r,\varphi ,z) = \sum_{j=1}^{N_{\rm pert}} v_{0,j}~{\rm cos}\left(k_{j}z+m_{j}\varphi + \phi_{j}\right)\\
\\
\times {\rm exp}\left[-\dfrac{(r-\Rs)^2}{2\sigma_{\rm pert}^2}\right]{\rm cos}\left(\varphi\right)
\end{array},
\ee
\be 
\label{eq:perty}
\begin{array}{c}
v_{\rm y}^{\rm pert}(r,\varphi ,z) = \sum_{j=1}^{N_{\rm pert}} v_{0,j}~{\rm cos}\left(k_{j}z+m_{j}\varphi + \phi_{j}\right)\\
\\
\times {\rm exp}\left[-\dfrac{(r-\Rs)^2}{2\sigma_{\rm pert}^2}\right]{\rm sin}\left(\varphi\right)
\end{array}.
\ee

\smallskip
The velocity perturbations are localised on the stream-background interface, with a penetration depth set by the parameter $\sigma_{\rm pert}$, which is $\Rs/16$ in all of our simulations. To comply with periodic boundary conditions, all wavelengths are harmonics of the box length, $k_{j}=2\pi n_{j}$, where $n_{j}$ is an integer, corresponding to a wavelength $\lambda_{j}=1/n_{j}$. Each simulation includes all wavenumbers in the range $n_{j}=2-64$, corresponding to all available wavelengths in the range $\Rs/2 - 16\Rs$. Each wavenumber is also assigned a symmetry mode, represented by the index $m_{j}$ in \equs{pertx} and \equm{perty}. For each wavenumber we include both an $m=0$ and an $m=1$ mode, corresponding to axisymmetric and helical modes, respectively. This yields a total of $N_{\rm pert}=2\times 63=126$ modes per simulation. In one simulation, M2.0\_d100\_Hm (\tab{sims}), we included modes with $m=0,1,2,3,4$, yielding a total of $N_{\rm pert}=5\times 63=315$ modes. Each mode is then given a random phase $\phi_{j} \in [0,2\pi)$. Different realizations with different phases are found to be extremely similar (P18; M19). The amplitude of each mode, $v_{0,j}$, is identical, with the rms amplitude normalised to $0.01\cs$.

\subsection{Tracing the Two Fluids} 
\label{sec:tracer}

\smallskip
Following P18; M19 and \citet{Aung19}, we use a passive scalar field, $\psi(r,\varphi,z,t)$, to track the growth of the shear layer and the mixing of the two fluids. Initially, $\psi=1$ and $0$ in the stream and background respectively. During the simulation, $\psi$ is advected with the flow such that the density of stream-fluid in each cell is $\rhos=\psi\rho$, where $\rho$ is the total density in the cell. 

\smallskip
The volume-weighted average radial profile of the passive scalar is given by 
\be 
\label{eq:volume-averaged-colour}
\overline{\psi}(r,t) = \frac{\int_{-L/2}^{L/2}\int_{0}^{2\pi} \psi_{(r,\varphi ,z,t)} r~{\rm d\varphi \,dz}}{2\pi r L}.
\ee
{\no}This is used to define the edges of the shear layer around the stream interface, $r(\overline{\psi}=\epsilon)$ on the background side and $r(\overline{\psi}=1-\epsilon)$ on the stream side, where $\epsilon$ is an arbitrary threshold. We set $\epsilon=0.04$ following M19, though our results are not strongly dependent on this choice. The background-side thickness of the shear layer is then defined as 
\be 
\label{eq:hb_def}
\hb \equiv {\rm max_r}r(\overline{\psi}=\epsilon)-\Rs,
\ee
{\no}while the stream-side thickness is defined as 
\be 
\label{eq:hs_def}
\hs \equiv \Rs-{\rm min_r}r(\overline{\psi}=1-\epsilon).
\ee

\section{Simulation results} 
\label{sec:results} 


\smallskip
In this section we present the results of the numerical simulations described in \se{methods}. in \se{Morph}, we discuss the evolution of stream morphology and eventual stream disruption. In \se{decel} we discuss mass entrainment onto the stream through the turbulent mixing layer, and subsequent stream deceleration and energy dissipation. In \se{turb} we discuss turbulence and heating in the stream driven by the instability. In \se{emiss} we present results for the total cooling emissivity generated by the instability.

\subsection{Stream Morphology and Disruption}
\label{sec:Morph}

\begin{figure*}
\begin{center}
\includegraphics[trim={0.18cm 0.08cm 0.08cm 0.58cm}, clip, width =0.98 \textwidth]{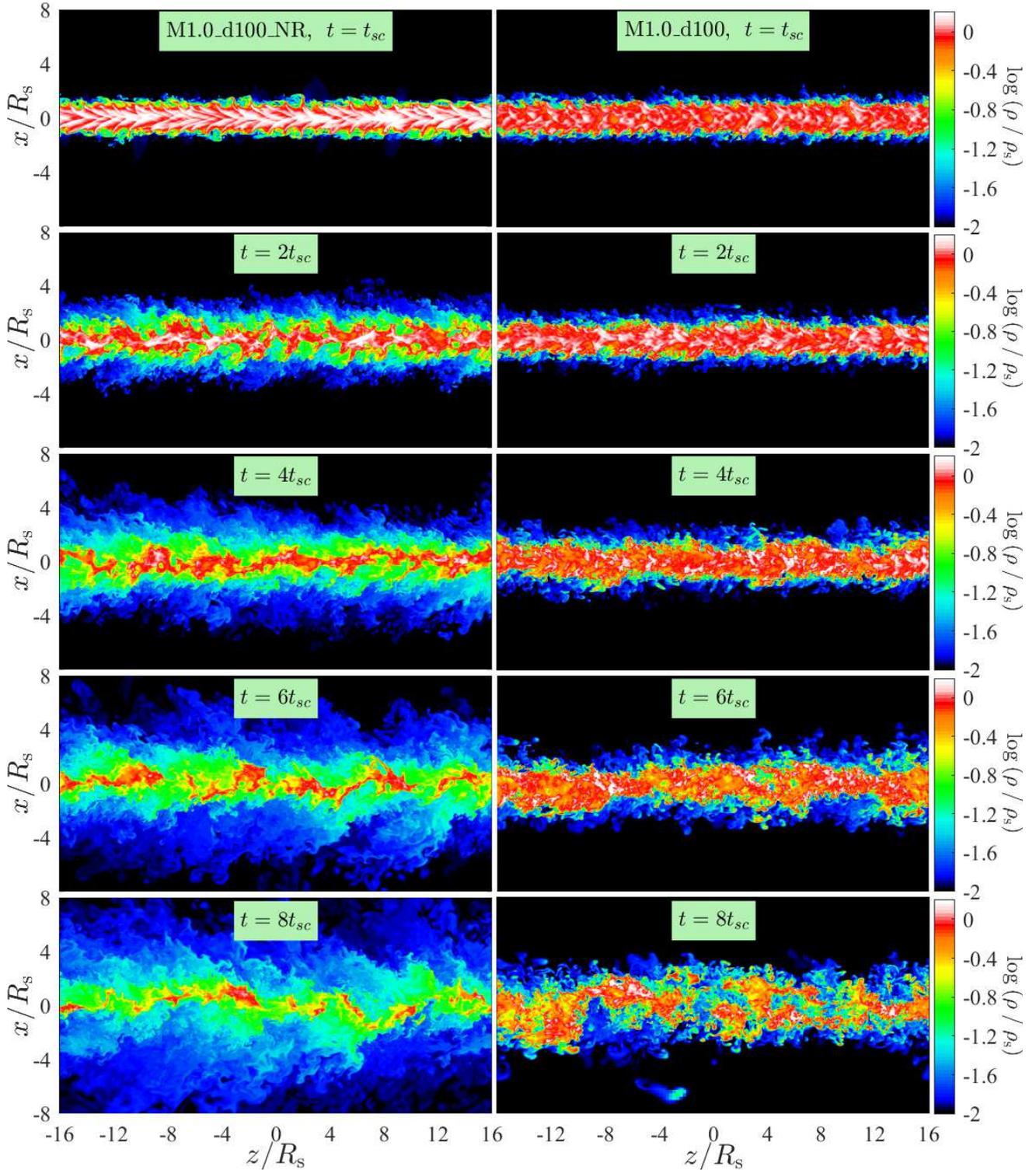}
\end{center}
\caption{Stream evolution without (left) and with (right) radiative cooling. Shown are snapshots of density normalized by the initial stream density, $\rho/\rho_{\rm s}$, in an infinitesimal slice through the $xz$ plane, an ``edge-on" view of the cylinder. The unperturbed initial conditions were $(\Mb,\delta)=(1,100)$, the left-hand column shows the NR simulation while the right-hand column shows the fiducial simulation (see \tab{sims}). The snapshot times in units of the stream sound crossing time, $\tsc=2\Rs/\cs$, are shown in each panel. While the density in the non-radiative case grows smaller as the stream expands into the background and is diluted, the density in the cooling simulation remains high until $t>8\tsc$. In this case, stream expansion is suppressed and the background condenses onto the stream since $\tcm<t_{\rm shear}$ (\tab{sims}). Eventually, the stream begins to break up due to thermal instabilities in the turbulent mixing layer, which lead to the formation of small dense clumps near the stream-background interface.}
\label{fig:density_panel_1} 
\end{figure*}

\begin{figure*}
\begin{center}
\includegraphics[trim={0.18cm 0.08cm 0.08cm 0.58cm}, clip, width =0.98 \textwidth]{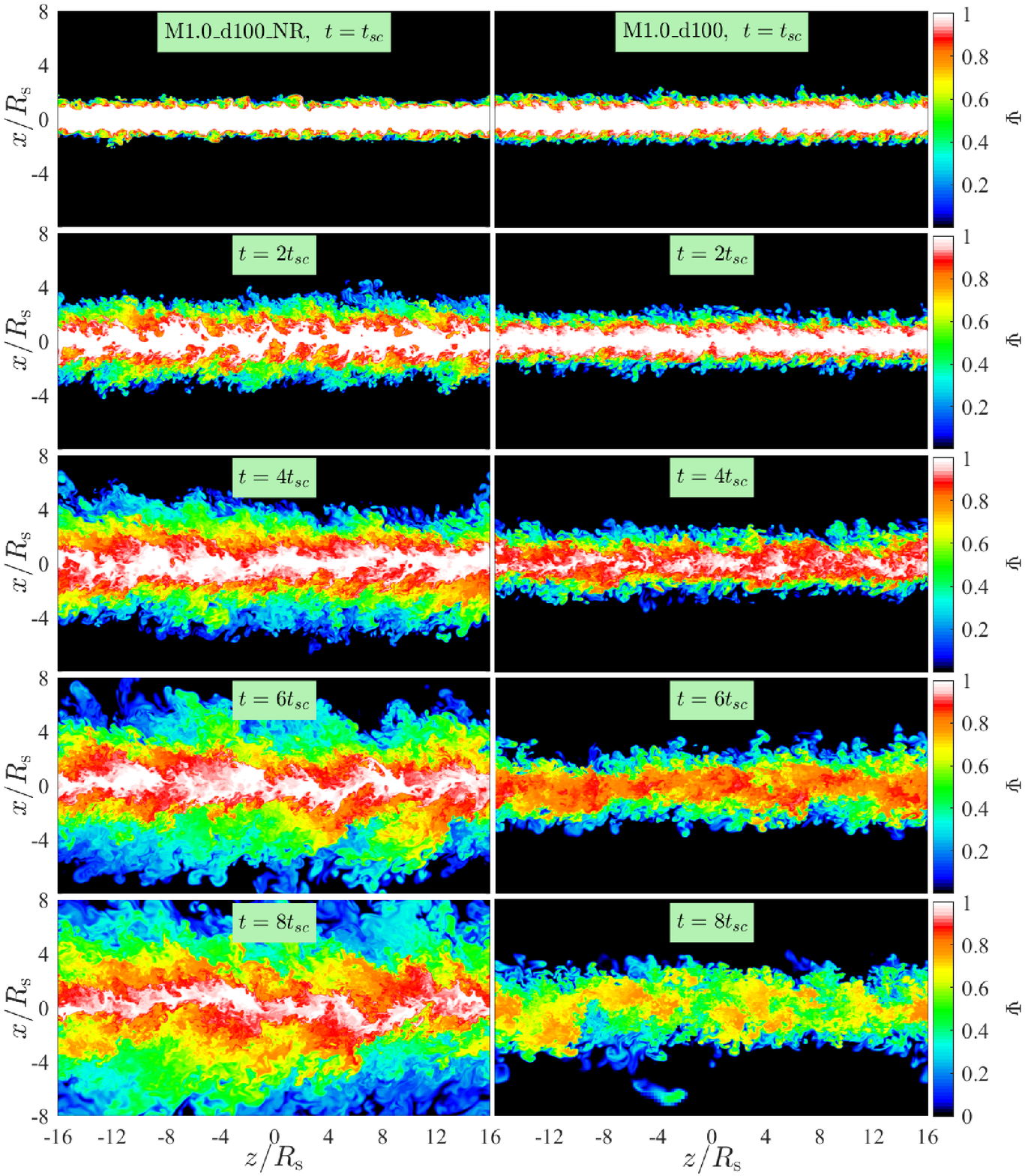}
\end{center}
\caption{Same as \fig{density_panel_1}, but showing the passive scalar, $\psi$ (\se{tracer}), rather than the  fluid density. In the non-radiative simulation, the stream expands into the background depositing stream fluid far from the initial interface. In the cooling simulation, on the other hand, the background condenses onto the stream and the mixing region is significantly smaller in volume, because $\tcm<t_{\rm shear}$. As a result, the fraction of mixed fluid near the stream centre, with $\psi<1$, increases much faster in the simulation with cooling.}
\label{fig:colour_panel_1} 
\end{figure*}

\smallskip
In \fig{density_panel_1} we show a time sequence of the density evolution for simulations M1.0\_d100\_NR (left) and M1.0\_d100 (right). We show a slice through the $y=0$ plane of the stream at times $t=(1,2,4,6,8)\tsc$. The differences between the radiative and non-radiative simulations are striking. At $t=\tsc$, the density in the non-radiative stream is larger because a shock was able to penetrate to the stream axis, while it was damped in the cooling simulation. However, this is a transient feature, likely influenced by the initial simulation setup. At all later times the average density in the non-radiative stream is smaller than in the radiative case. In the former, the stream expands into the background and is diluted. By $t=8\tsc$, there is hardly any gas left with density comparable to the initial stream density and the average density along the stream axis is $\sim 0.2\rhos$. On the other hand, the density in the cooling simulation remains high until late times, as stream expansion is suppressed and the background condenses onto the stream. This is expected, since in this case $\tcm<t_{\rm shear}$ (\tab{sims}). At $t=8\tsc$ the stream has begun to break up, forming small dense clumps near the stream-background interface. 
These may be due to thermal instabilities in the mixing region, which is pressure confined by the external gas outside this region, leading to fragmentation to the cooling length, $l_{\rm cool}=\cs t_{\rm cool}$ \citep{McCourt18}. They may also be due to secondary instabilities following the growth of long wavelength modes, seen at $t=6$ and $8\tsc$, similar to non-linear disruption by body modes seen in M19. This long wavelength mode would have time to manifest in the radiatively cooling simulation, because stream disruption by short-wavelength surface modes is suppressed. A detailed study of the origin of these clumps, and their mass/size distribution, is beyond the scope of this paper and left for future work. For our current purposes, suffice to say that the stream density remains high and that it maintains a relatively collimated structure.

\smallskip
In \fig{colour_panel_1}, we show the evolution of the passive scalar, $\psi$, in the same snapshots shown in \fig{density_panel_1}. In the non-radiative case, the stream expands into the background and stream fluid is found very far from the stream axis. At $t=8\tsc$, we have $\hb\sim 8.5\Rs$, as defined in \equ{hb_def}. On the other hand, there is still a core of unmixed stream fluid with $\psi\lsim 1$ near the stream axis, consistent with the fact that for $(\Mb,\delta)=(1.0,100)$, $t_{\rm dis}\sim 11\tsc$ (\equnp{tdis_non_rad}). The situation is markedly different in the radiative case. Since $\tcm<t_{\rm shear}$, stream expansion is suppressed and $\hb$ is only $\sim 3\Rs$ at $t=8\tsc$. However, background material condenses onto the stream and mixes with it, lowering $\psi$ near the stream axis much faster than in the non-radiative case. By $t\sim 3\tsc$, $\hs=\Rs$. Based on the definition of stream disruption used in P18, M19, and \citet{Aung19}, this would imply that the stream is disrupted at $t\sim 3\tsc$, nearly 4 times faster than the non-radiative case. However, this is clearly not the case as evidenced by the large densities and collimated structure present in the stream until very late times (\fig{density_panel_1}). This highlights the need for an alternative definition of stream disruption. 

\begin{figure*}
\begin{center}
\includegraphics[trim={0.18cm 0.0cm 0.03cm 0.0cm}, clip, width =0.98 \textwidth]{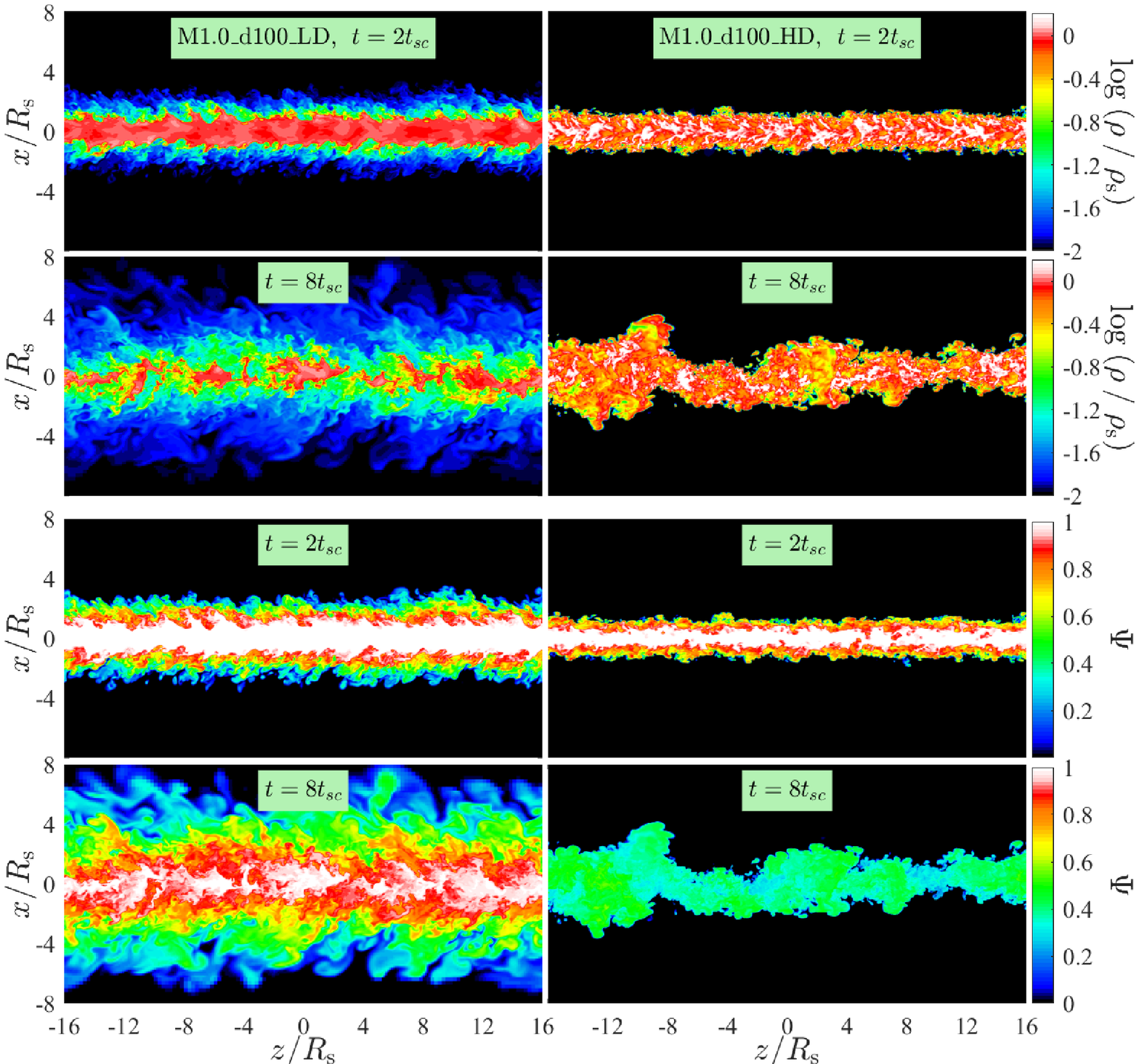}
\end{center}
\caption{Similar to \figs{density_panel_1}-\figss{colour_panel_1}, but for the LD (left) and HD (right) simulations with $(\Mb,\delta)=(1.0,100)$ (see \tab{sims}). The top two rows show the density at $t=2\tsc$ and $8\tsc$, while the bottom two rows show the passive scalar at the same times. Since the LD simulation has $\tcm>t_{\rm shear}$, its behaviour is qualitatively similar to the non-radiative (NR) case shown in \figs{density_panel_1}-\figss{colour_panel_1}. In the HD simulation, on the other hand, $\tcm/t_{\rm shear}$ is much smaller than in the fiducial case shown in \figs{density_panel_1}-\figss{colour_panel_1}. This leads to a narrower turbulent mixing layer, more condensation of background material onto the stream, higher densities and more mixed fluid near the stream centre.}
\label{fig:panels_2} 
\end{figure*}

\smallskip
In \fig{panels_2} we show slices of the density and the passive scalar at $t=2\tsc$ and $t=8\tsc$ for the simulations M1.0\_d100\_LD (left) and M1.0\_d100\_HD (right). In the former, $\tcm>t_{\rm shear}$, while in the latter $\tcm<<t_{\rm shear}$ (\tab{sims}). Based on the arguments outlined in \se{theory_cool}, the LD case should thus be qualitatively similar to the NR case, while the HD case should be similar to the fiducial simulation. This is indeed the case. 
In the LD simulation, the stream expands into the background and its density is diluted. This occurs at a slightly slower rate than in the NR case, with $\hb\sim 6.5\Rs$ at $t=8\tsc$, and the average density along the stream axis only half the initial value. This is consistent with simulations of the evolution of cold clouds traveling in a hot background, which find that cooling can delay cloud disruption by a factor of a few when $\tcm>t_{\rm cc}$, but it does not qualitatively alter the evolution (\citealp{Scannapieco15,Schneider17}; G18). On the other hand, in the HD simulation, the density remains very high and the stream remains highly collimated even at $t>8\tsc$. Likewise, the passive scalar reaches even lower values than in the fiducial case near the stream axis, since the flow of background material through the turbulent mixing region onto the stream is more efficient (\equnp{vmix_gronke}). Stream breakup into dense clumps appears suppressed compared to the fiducial simulation (bottom right panel of \fig{density_panel_1}). If such breakup is due to secondary instabilities following the growth of long wavelength modes, these may be stabilized by the more efficient cooling. On the other hand, if the breakup is due to thermal shattering of a pressure confined medium, the clumps may simply be unresolved in this case because the cooling length in the mixing layer is of the order of a single cell. In this case, we suspect that simulations that resolve the cooling length would produce a similar fragmented structure to that observed in the fiducial case (see \citealp{M19b}). As stated above, the origin and evolution of these small clumps is beyond the scope of the current paper. 

\subsection{Mass Entrainment, Stream Deceleration, and Loss of Kinetic and Thermal Energy}
\label{sec:decel}

\begin{figure*}
\begin{center}
\includegraphics[trim={0.05cm 0.04cm 0.0cm 0.02cm}, clip, width =0.98 \textwidth]{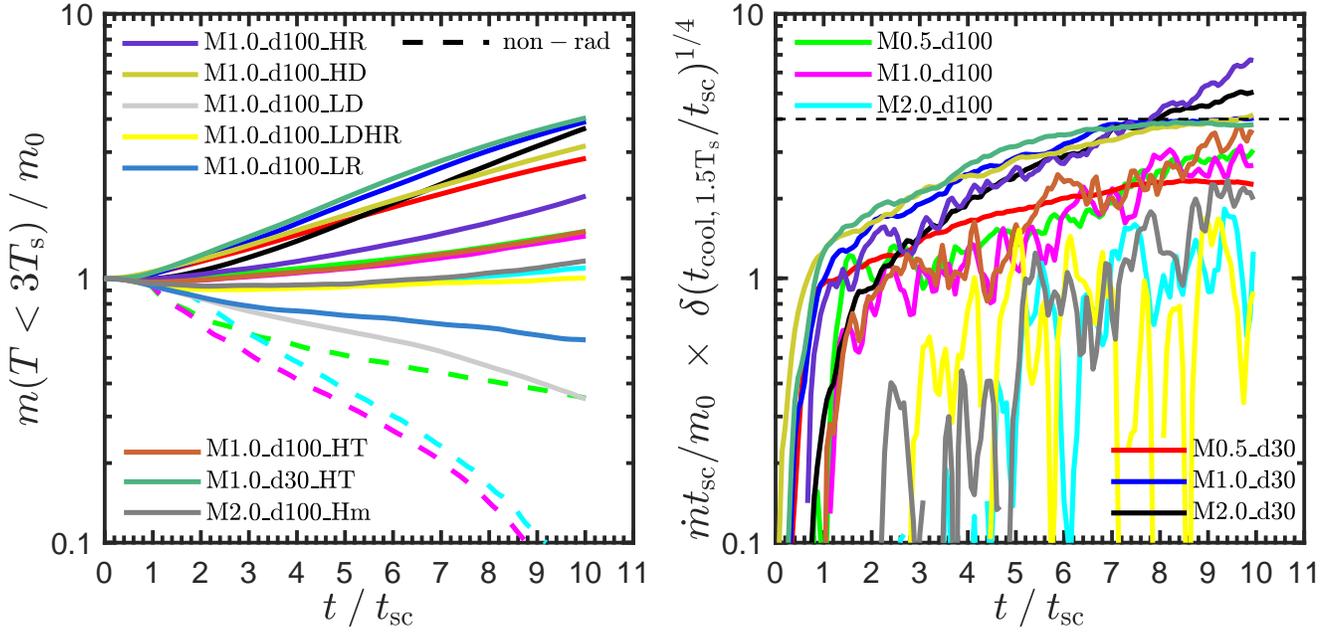}
\end{center}
\caption{Evolution of cold gas mass in the simulations. \textit{Left:} Mass of gas with $T<3\Ts$ as a function of time, normalized to the initial stream mass. Different colored lines show different simulations (see \tab{sims}), and the dashed lines show the three non-radiative (NR) simulations. In cases where $\tcm>t_{\rm shear}$, the two LD and three NR simulations, the cold mass decreases with time as the stream expands into the background and is diluted. In all other simulations, where $\tcm<t_{\rm shear}$, the cold mass increases as background mass cools and condenses onto the stream through the mixing layer. In many cases, the stream mass can increase by a factor of $\sim (3-4)$ within $\sim 10\tsc$. \textit{Right:} Entrainment rate of background mass onto the stream, defined as the time derivative of the curves in the left-hand panel, for all cases where $\tcm<t_{\rm shear}$. These are then scaled according to \equ{mdot_gronke}.  When properly scaled, the entrainment rates are within $\lsim 50\%$ of each other at all times, and the assymptotic value is roughly $\sim 4$ (horizontal dashed line) as predicted. In cases where $\tcm\gsim 0.15 t_{\rm shear}$ (M\_2.0\_d100, M\_2.0\_d100\_Hm, and M\_1.0\_d100\_LDHR), the entrainment rates are slightly lower than predicted.
}
\label{fig:mass_panels} 
\end{figure*}

\smallskip
In \fig{mass_panels} we address the cooling and entrainment of background mass onto the stream. In the left hand panel we show the evolution of cold mass in all our simulations. We define ``cold" as gas with temperature within a factor 3 of the initial stream temperature, $T<3\Ts$ (see \tab{sims} for $\Ts$ values), though our results are not sensitive to this precise choice. 
In simulations where $\tcm/t_{\rm shear}\lsim 0.1$, the cold mass begins increasing within less than one stream sound crossing time from the start of the simulation, and grows monotonically as background mass cools and condenses onto the stream. In many cases, the cold mass at $t\sim 10\tsc$ is $\sim (3-4)$ times larger than the initial stream mass. In simulations where $0.1<\tcm/t_{\rm shear}< 1$ (M\_2.0\_d100, M\_2.0\_d100\_Hm, M\_1.0\_d100\_LDHR), the cold mass decreases at first, but begins growing after $\sim (2-3)\tsc$, and by the end of the simulation at $10\tsc$ there is more cold gas than at the start of the simulation. However, in simulations where $1\lsim \tcm/t_{\rm shear}$ (M\_1.0\_d100\_LD, M\_1.0\_d100\_LR), the cold mass decreases with time, since cooling is not fast enough to prevent stream shredding and dilution. This is qualitatively consistent with the NR simulations (dashed lines in the figure), though the cooling does slow down the rate of mass loss and stream expansion, as discussed above. Indeed, the cold mass declines slower for $\tcm/t_{\rm shear}\sim 0.97$ (M\_1.0\_d100\_LR) than for $\tcm/t_{\rm shear}\sim 2.44$ (M\_1.0\_d100\_LD).

\smallskip
In the right hand panel of \fig{mass_panels} we show the entrainment rate of background gas onto the stream in the simulations, defined as the time derivative of the curves in the left-hand panel. Note that the LD, SR, and NR simulations, where the cold mass is decreasing, are not shown. We compare the measured entrainment rates to the prediction from \equ{mdot_gronke}, by scaling them accordingly. When properly scaled, all cases where $\tcm/t_{\rm shear}\lsim 0.1$ coincide to within $\sim 50\%$ at all times. The assymptotic value is $\sim 4$, in agreement with \equ{mdot_gronke}. We obtain very similar results by replacing $t_{\rm cool,1.5\Ts}$ in \equ{mdot_gronke} with the cooling time evaluated in the temperature range $(1.2-2)\Ts$. The three simulations where $0.1<\tcm/t_{\rm shear}< 1$ have lower entrainment rates than predicted by \equ{mdot_gronke} by a factor of $\sim 2-4$. However, it is possible that these simulations have not yet reached steady-state entrainment, and that this discrepancy will decrease at later times. Additionally, the entrainment rate in the M1.0\_d100\_HR simulation, with $\Rs=6\kpc$, seems not to have converged by $t=10\tsc$. The reasons for this are not entirely clear, and we expect that the entrainment rate will likely converge at a value $\lsim 2$ times larger than the prediction at $t\gsim 10\tsc$. As the entrainment values are still in reasonable agreement with our model compared to other simulations, we do not currently address this issue further.

\begin{figure*}
\begin{center}
\includegraphics[trim={0.05cm 0.04cm 0.035cm 0.02cm}, clip, width =0.98 \textwidth]{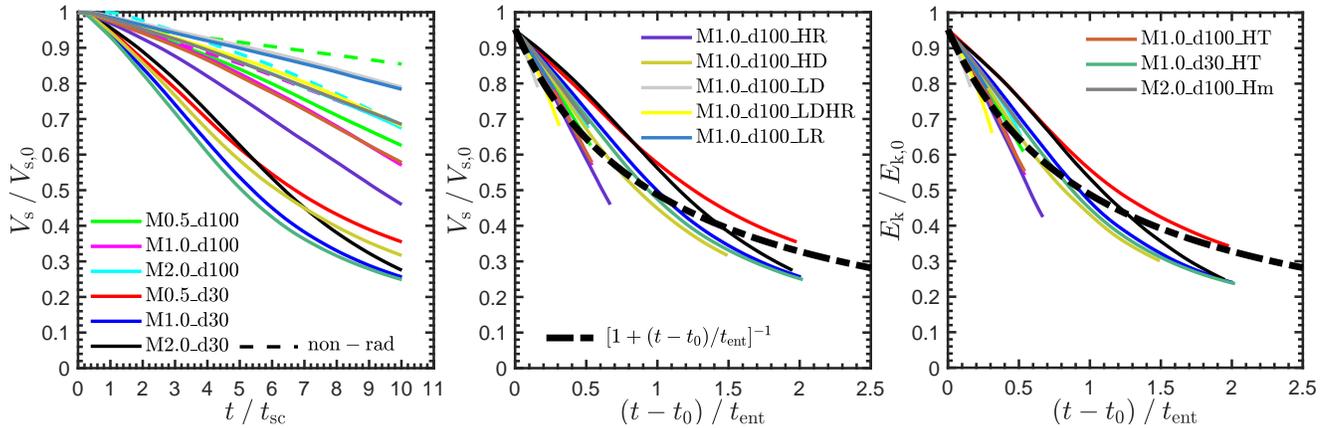}
\end{center}
\caption{Stream deceleration and kinetic energy loss. Line styles and colours are as in \fig{mass_panels}.
\textit{Left:} Centre of mass velocity of stream fluid (weighted by the passive scalar $\psi$) normalized by its initial value, as a function of time normalized by the stream
sound crossing time. By comparing the solid magenta, yellow and grey lines, in is clear that deceleration in radiatively cooling streams does not depend only on $(\Mb,\delta)$, but on $t_{\rm cool}$ as well. \textit{Centre:} Same as the left-hand panel, but with the time axis normalized to $t_{\rm ent}$ (\equnp{mass_rad_approx}), and the zero-point shifted to $t_0$, the time at which the velocity reaches $95\%$ of its initial value, in order to avoid focusing on the initial transient before a steady-state is reached. The thick dot-dashed curve shows the predicted deceleration rate (\equnp{decel_rad}), which is found to be a good match to all radiative simulations. \textit{Right:} Total kinetic energy associated with laminar flow of both stream and background fluid, normalized by its initial value, as a function of time normalized by $t_{\rm ent}$. The kinetic energy loss is well described by \equ{Ek_rad}, marked by the thick dot-dashed line.}
\label{fig:decel_panels} 
\end{figure*}

\smallskip
As background mass is entrained onto the stream, the stream decelerates due to momentum conservation.
In the left hand panel of \fig{decel_panels}, we show the stream velocity, $V_{\rm s}$, normalized by its initial value as a function of time normalized by the sound crossing time. $V_{\rm s}$ is computed as the average velocity in the ${\hat {z}}$ direction, weighted by the stream mass\footnote{We weight by stream mass for consistency with P18, M19, and \citet{Aung19}. However, the results are nearly identical when using total mass, since the stream and the background are so well mixed, as evidenced by the uniformity of $\psi$ seen in \figs{colour_panel_1} and \figss{panels_2}.} in each cell, $m_{\rm s}=\psi m$, where $m$ is the total cell mass. It is instructive to compare the magenta, light-grey, and gold lines, which refer to simulations with $(\Mb,\delta)=(1.0,100)$, but different stream densities and thus different cooling times. These three simulations have very different deceleration rates, showing that \equs{stream_deceleration}-\equm{tau_dec}, which were found to describe stream deceleration in the non-radiative case (M19), are not a valid description for the deceleration of radiatively cooling streams. Note also that the LD simulation (light-grey line) decelerates slower than the corresponding non-radiative case (dashed magenta line), because cooling slows stream expansion in the former despite it being in the regime where $\tcm>t_{\rm shear}$. 

\smallskip
Conservation of momentum yields $V(t)=V_0 m_0/m(t)$, where $m$ represents the cold mass flowing with the stream.\footnote{For $\Mb\gsim 1$, some momentum is transferred to the hot gas far from the stream. However, this turns out to be very small and does not affect the arguments presented in the text.} As an approximation, we assume 
\be 
\label{eq:mass_rad_approx}
m(t) = m_0+\left<{\dot {m}}\right>t = m_0\left(1+\dfrac{t}{t_{\rm ent}}\right),
\ee
{\no}where $\left<{\dot {m}}\right>$ is the average entrainment rate and we have introduced the entrainment timescale, 
\be 
\label{eq:tent}
t_{\rm ent}\equiv\frac{m_0}{\left<{\dot {m}}\right>}\sim \frac{\delta}{2}\left(\frac{t_{\rm cool,1.5\Ts}}{\tsc}\right)^{1/4} \tsc, 
\ee 
{\no}motivated by the right-hand panel of \fig{mass_panels}. This results in a predicted velocity as a function of time 
\be 
\label{eq:decel_rad}
V_{\rm s}(t) = \frac{V_{\rm s,0}}{1+t/t_{\rm ent}}.
\ee
{\no}In the center panel of \fig{decel_panels}, we show the stream velocity as a function of time normalized to $t_{\rm ent}$. All simulations now behave very similarly, and are 
well described by \equ{decel_rad}, shown by the thick dot-dashed line. Interestingly, even the LD and SR simulations, where $\tcm\gsim t_{\rm shear}$ and the cold mass declines with time (\fig{mass_panels}), seem to follow this relation.

\smallskip
As discussed in \se{theory_khi}, since the deceleration is caused by mass entrainment and momentum conservation, the kinetic energy associated with the bulk laminar flow declines as $E_{\rm k}(t)/E_{\rm k,0}=V(t)/V_0$, yielding
\be 
\label{eq:Ek_rad}
E_{\rm k}(t) = \frac{E_{\rm k,0}}{1+t/t_{\rm ent}}.
\ee
{\no}In the right-hand panel of \fig{decel_panels}, we show the total kinetic energy associated with the laminar flow of both stream and background fluid as a function of time normalized to $t_{\rm ent}$. To compute the kinetic energy, we first compute the mass-weighted mean velocity in the ${\hat{z}}$ direction at each radius $r$, 
\be 
\label{eq:vzr}
\widetilde{v_{\rm z}}(r)\equiv \overline{\rho v_{\rm z}}(r)/\overline{\rho}(r), 
\ee 
{\no}where $\overline{\rho}(r)$ and $\overline{\rho v_{\rm z}}(r)$ are obtained analogously to \equ{volume-averaged-colour}. We then compute the kinetic energy associated with this laminar flow, by integrating $0.5 \rho \widetilde{v_{\rm z}}^2$ over the full simulation volume. As expected, \equ{Ek_rad} (thick dot-dashed line) is a good match to the simulation data. 

\begin{figure} 
\begin{center}
\includegraphics[trim={0.03cm 0.04cm 0.0cm 0.02cm}, clip, width =0.49 \textwidth]{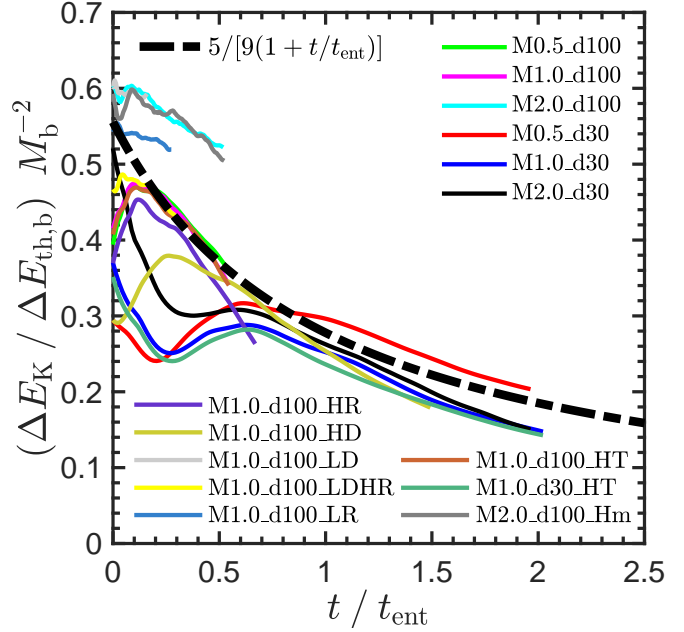}
\end{center}
\caption{Ratio of kinetic to thermal energy loss in the simulations. We show this ratio, normalized by $\Mb^2$, as a function of time normalized by $t_{\rm ent}$. The thick dot-dashed line shows \equ{energy_ratio}, which is a good match to all simulations at $t\gsim 0.5t_{\rm ent}$.}
\label{fig:EK_over_Eth} 
\end{figure}

\smallskip
In addition to the stream losing kinetic energy, the background gas entrained by the stream loses thermal energy, and we wish to examine which dominates the overall energy loss. The rate of thermal energy loss is 
\be 
\label{eq:erad}
{\dot{E}}_{\rm th,b} = {\dot {m}}(e_{\rm b}-e_{\rm s}) \simeq {\dot {m}}e_{\rm b} = \frac{{\dot {m}}\cb^2}{\gamma(\gamma-1)} = \frac{9 m_0 \cb^2}{10t_{\rm ent}},
\ee 
{\no}where $e_{\rm b}=P/[(\gamma-1)\rhob]$ is the thermal energy per unit mass of the background fluid which is larger than that in the cold component by a factor $\delta>>1$, $\gamma=5/3$ is the adiabatic index of the gas, and $\cb^2=\gamma P/\rhob$ is the adiabatic sound speed in the background. In the final equation we have used \equ{mass_rad_approx} to approximate ${\dot{m}}$. The total thermal energy lost by time $t$ is thus 
\be 
\label{Delta_eth}
\Delta E_{\rm th,b}=\frac{9 m_0 \cb^2 t}{10 t_{\rm ent}}.
\ee
{\no}Inserting $E_{\rm k,0}=0.5m_0 V_0^2 = 0.5 m_0 \Mb^2 \cb^2$ into \equ{Ek_rad}, we obtain the total kinetic energy lost by time $t$,
\be 
\label{Delta_ek}
\Delta E_{\rm k}=\frac{m_0 \Mb^2 \cb^2}{2}\frac{t/t_{\rm ent}}{1+t/t_{\rm ent}}.
\ee
{\no}We thus obtain the ratio of kinetic to thermal energy loss as a function of time, 
\be 
\label{eq:energy_ratio}
\frac{\Delta E_{\rm k}}{\Delta E_{\rm th,b}} \sim \frac{5\Mb^2}{9\left(1+t/t_{\rm ent}\right)}.
\ee
{\no}\Fig{EK_over_Eth} shows this ratio normalized by $\Mb^2$ as a function of time normalized by $t_{\rm ent}$. $E_{\rm th,b}$ is computed in the simulations as the integral of $\rhob e_{\rm b}=(1-\psi)\rho e_{\rm b}$ over the full simulation volume. \Equ{energy_ratio} is a good approximation to all simulations, especially at times $t\gsim 0.5t_{\rm ent}$. In particular, this means that most of the energy radiated away is the thermal energy of the background gas, rather than the stream kinetic energy. 

\begin{figure*}
\begin{center}
\includegraphics[trim={0.06cm 0.04cm 0.03cm 0.02cm}, clip, width =0.98 \textwidth]{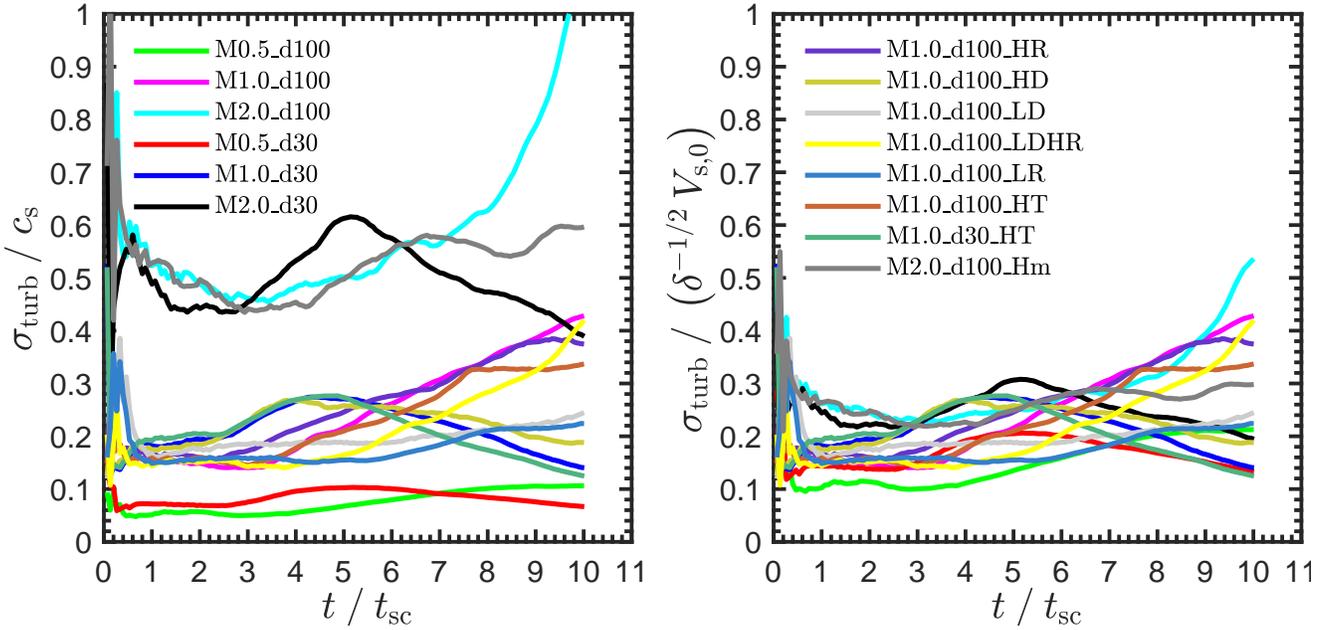}
\end{center}
\caption{Turbulence in the mixing zone as a function of time. \textit{Left:} The turbulent velocity normalized by the initial stream sound speed. The instability drives subsonic turbulence within the stream, with turbulent Mach numbers in the range $\sim (0.1-0.5)$. \textit{Right:} The turbulent velocity normalized by $\delta^{-1/2}\,V_{\rm s,0}$. In all our simulations, $\sigma_{\rm turb}\sim 0.2\,\delta^{-1/2}\,V_{\rm s,0}$.
}
\label{fig:turb_panels} 
\end{figure*}

\begin{figure}
\begin{center}
\includegraphics[trim={0.02cm 0.04cm 0.03cm 0.02cm}, clip, width =0.48 \textwidth]{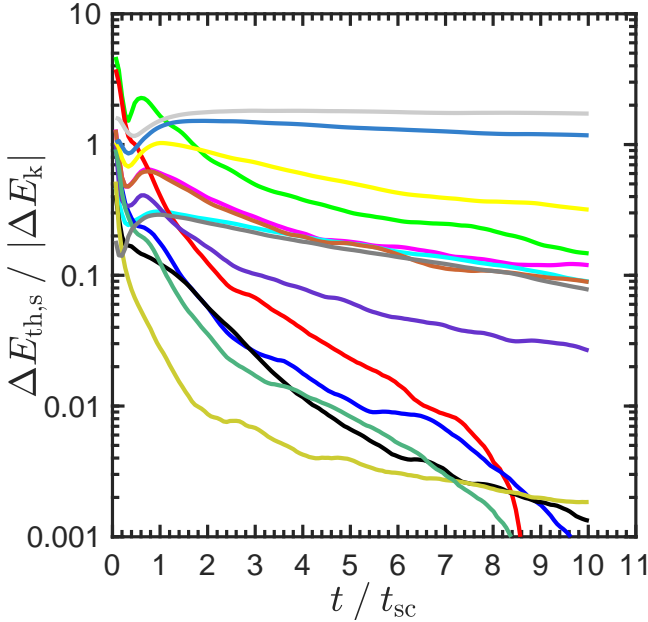}
\end{center}
\caption{Thermal energy increase in the stream from time 0 to $t$, normalized by the laminar kinetic energy lost during this time. This ratio decreases with time and with $\tcm/t_{\rm shear}$ (\tab{sims}), is rarely larger than $10\%$, and is often less than $1\%$. The legend is omitted for clarity, but is identical to previous figures.
}
\label{fig:Eths}
\end{figure}

\subsection{Turbulence, Heating, and Mixing}
\label{sec:turb}

\smallskip
Prior to being radiated away, some fraction of the energy lost from the system as described above will be converted into turbulent kinetic energy and thermal energy in the stream. In this subsection we estimate the fraction of the energy budget in each of these components over long timescales. We also examine the density and temperature distributions in the stream and the mixing of stream and background material. 

\subsubsection{Turbulent Energy and Mach Number}

\smallskip
We evaluate the magnitude of turbulence in \fig{turb_panels}. 
In the left-hand panel, we show the ratio of turbulent velocity within the mixing zone, $\sigma_{\rm turb}$, to the initial sound speed in the stream, $\cs$. As we show below, the mean stream temperature does not vary much, so this ratio is roughly the turbulent Mach number within the mixing zone. $\sigma_{\rm turb}$ is measured in the simulations as
\be 
\label{eq:sig_turb_def}
\sigma_{\rm turb}^2 = \dfrac{\int_{\Rs-\hs}^{\Rs+\hb} \overline{\rho}_{(r)} \left[(v_{\rm z}-\widetilde{v_{\rm z}} )^2 + v_{\rm x}^2+v_{\rm y}^2\right] r~ {\rm dr}}{\int_{\Rs-\hs}^{\Rs+\hb} \overline{\rho}\, r ~{\rm dr}}.
\ee 
{\no}Note that $\hs=\Rs$ after $\sim (1-3)\tsc$ in all simulations with $\tcm<t_{\rm shear}$. The turbulene generated by the instability is subsonic, with turbulent Mach numbers in the range $\sim (0.1-0.5)$, or $\sigma_{\rm turb}\sim (2-10)\kms$. This is of order the velocity of material flowing through the mixing layer (\equnp{vmix_gronke}), and is less than the turbulent velocities in the non-radiative case (M19). 

\smallskip
In the right-hand panel of \fig{turb_panels} we show the turbulent velocity normalized by $\delta^{-1/2}\,V_{\rm s,0}$. This reduces the scatter between simulations significantly, and we find $\sigma_{\rm turb}\sim 0.2\,\delta^{-1/2}\,V_{\rm s,0}$. This shows that $\rhos \sigma_{\rm turb}^2\propto \rhob V_{\rm s,0}^2$, which in the rest frame of the stream implies that the turbulent energy induced in the stream is proportional to the kinetic energy of the shearing motion. For $\delta\sim (30-100)$, and for a total mass within the mixing layer $\sim (1-4)$ times the initial stream mass (\fig{mass_panels}), the turbulent kinetic energy is $\sim (0.5-5) \times 10^{-3}$ times the initial kinetic energy of the stream. This is completely negligible compared to the lost laminar kinetic energy (\fig{decel_panels}) or background thermal energy (\fig{Eths}).

\subsubsection{Heating of the Stream}

\smallskip
In \fig{Eths} we show the increase in the thermal energy of the stream fluid from time 0 to time $t$, normalized by the laminar kinetic energy lost during this time interval.\footnote{As with $E_{\rm turb}$, we are not claiming that the source for the increase in $E_{\rm th,s}$ is exclusively $\Delta E_{\rm k}$ as opposed to $\Delta E_{\rm th,b}$, but rather only using $\Delta E_{\rm k}$ as a means for quantifying $\Delta E_{\rm th,s}$.} $E_{\rm th,s}$ is computed by integrating $\rhos e_{\rm s}=\psi \rho e_{\rm s}$ over the simulation volume. The fraction of lost kinetic energy that is converted to thermal energy in the stream decreases with time and with $\tcm/t_{\rm shear}$ (see \tab{sims}). The increase in $E_{\rm th,s}$ is due both to turbulent decay and to mixing with hot background fluid, which can occur at lower densities farther from the stream axis as the ratio $\tcm/t_{\rm shear}$ increases (\fig{panels_2}). For the M1.0\_d100\_LD and M1.0\_d100\_LR simulations, where $\tcm>t_{\rm shear}$, the increase in thermal energy of stream fluid compensates for roughly all the lost kinetic energy. For the other cases, the fraction of lost kinetic energy maintained as thermal energy decreases with time, saturating at $\sim 10\%$ for cases where $\tcm\sim 0.1t_{\rm shear}$, and decreasing to $<1\%$ for cases where $\tcm\lsim 0.01t_{\rm shear}$.
\subsubsection{Temperature and Density Distribution in the Stream}

\begin{figure}
\begin{center}
\includegraphics[trim={0.01cm 0.0cm 0.02cm 0.00cm}, clip, width =0.495 \textwidth]{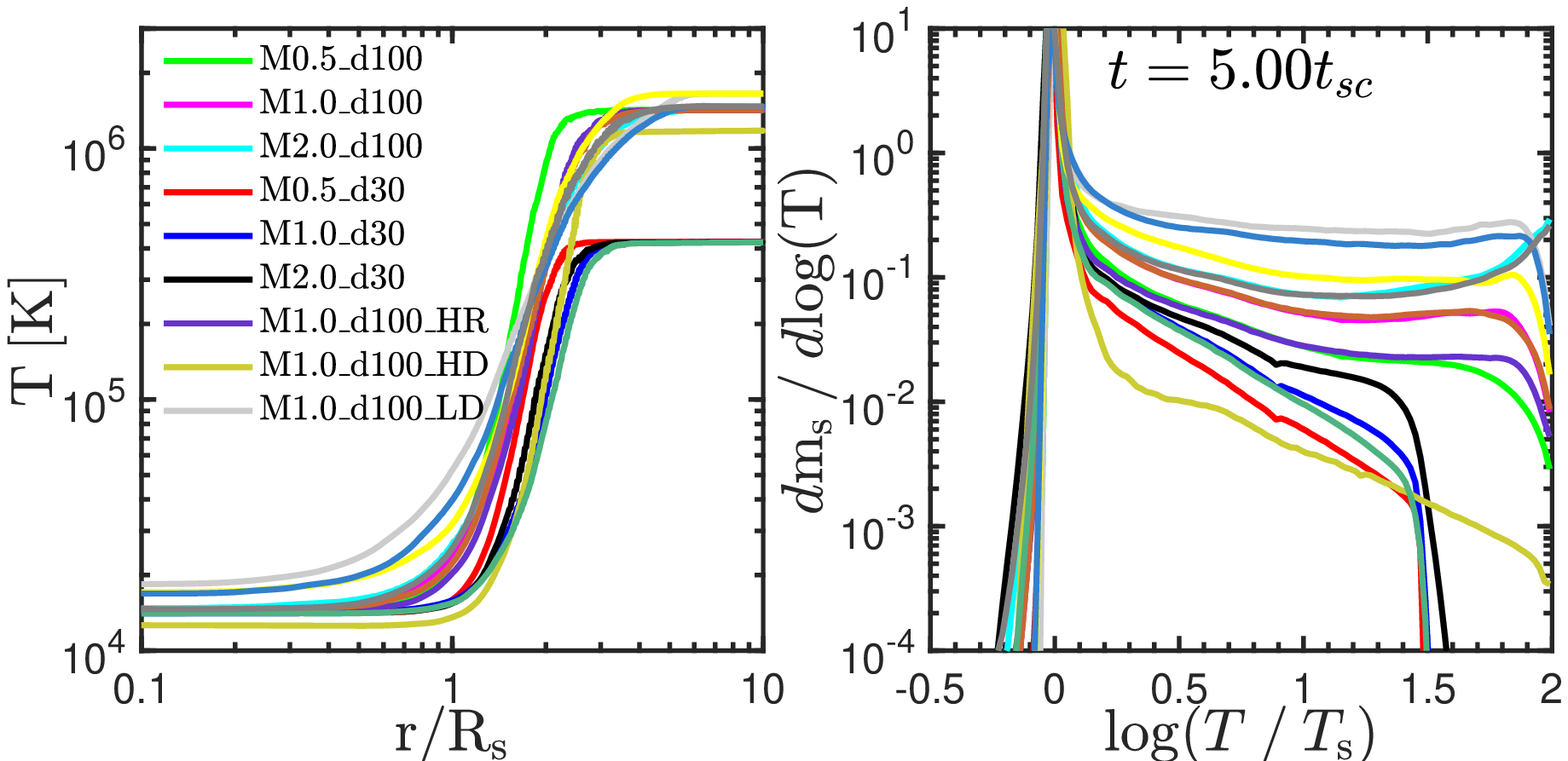}
\includegraphics[trim={0.01cm 0.0cm 0.02cm 0.00cm}, clip, width =0.495 \textwidth]{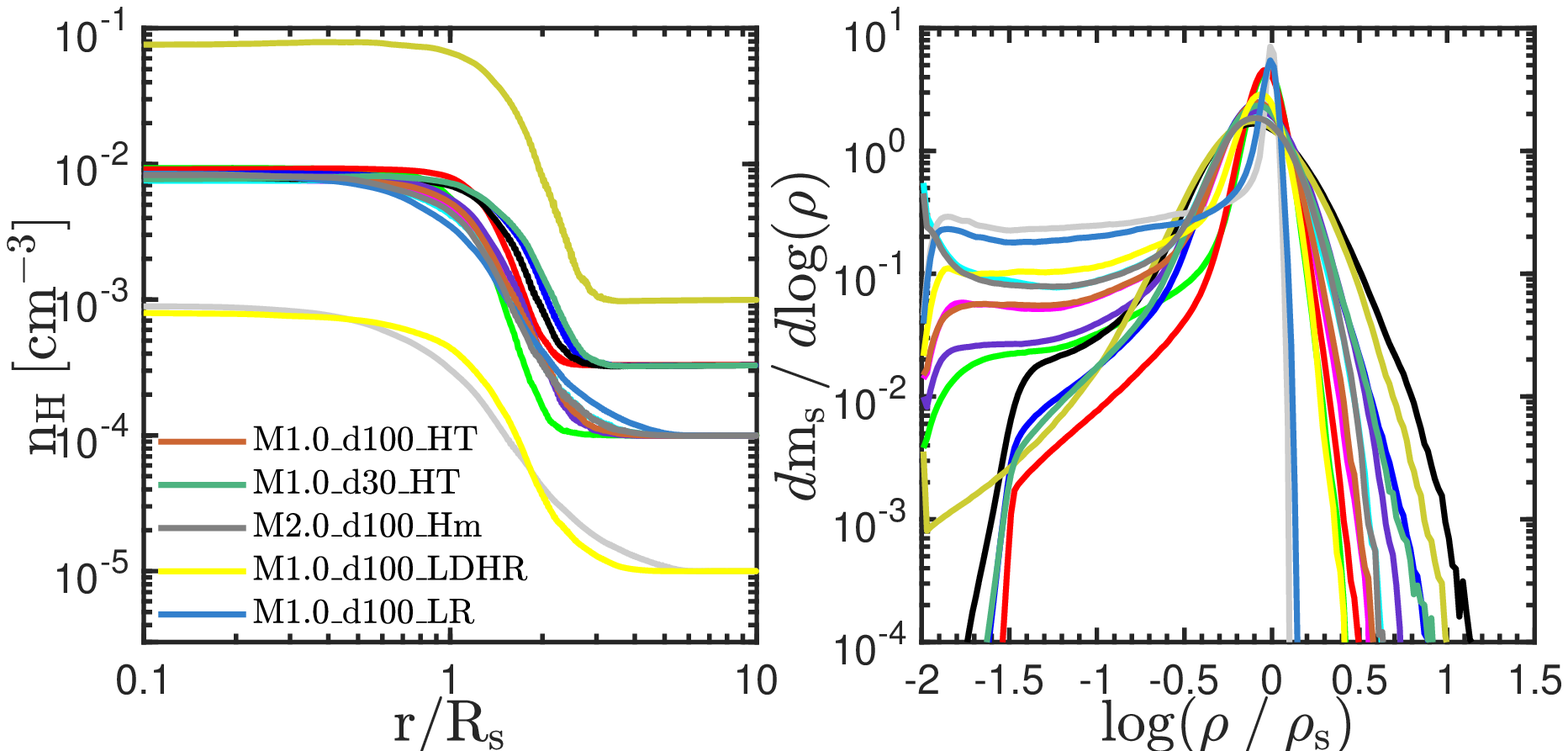}
\end{center}
\caption{Temperature \textit{(top panels)} and density \textit{(bottom panels)} distributions in the stream at $t=5\tsc$. \textit{Left panels:} Radial temperature and Hydrogen number density profiles as a function of $r/R_{\rm s}$. All profiles appear qualitatively similar at all times $t>\tsc$. The transition between the stream and background values begins at $r\lsim \Rs$, while its outer radius increases with time. \textit{Right panels:} Temperature and density PDFs weighted by stream-mass, $m_{\rm s}=\psi m$. Temperatures and densities have been normalized by the initial stream values, $T_{\rm s}$ and $\rho_{\rm s}$, respectively. The PDFs at all times $t\gsim \tsc$ are very similar. The temperature PDFs are sharply peaked at $T_{\rm s}$, with very little mass at $T<T_{\rm s}$ and a distribution towards higher $T$ which increases with the ratio $\tcm/t_{\rm shear}$. The density PDFs tend to peak at densities $\lsim 0.1~{\rm dex}$ lower than $\rhos$, and resemble a log-normal distribution around the peak. Large overdensities are apparent in all simulations with $\tcm<t_{\rm shear}$, reaching up to a factor of $\sim 10$ in the most rapidly cooling simulations. 
}
\label{fig:temp_dens_panels} 
\end{figure}

\smallskip
In \fig{temp_dens_panels} we examine the temperature (top panels) and density (bottom panels) distributions of stream gas at $t=5\tsc$ for all simulations. 
In the top-left-hand panel we show the radial temperature profiles, weighted by total mass in each radial bin. The profiles look qualitatively similar at all times $t\gsim \tsc$. The transition from $\Ts$ to $\Tb$ occurs in the radial range $0.8\lsim (r/\Rs)\lsim 2$, though the radius where $T$ reaches $\Tb$ increases with time. In the top-right-hand panel we show the temperature PDFs weighted by stream-mass, $m_{\rm s}=\psi m$. The PDFs are very similar at all times $t>\tsc$. They are sharply peaked near the initial stream temperature, $\Ts$, with very little stream mass at lower temperatures. The lowest temperatures reached in the simulations are $\gsim 0.6\Ts$. The width of the distribution towards higher temperatures increases with $\tcm/t_{\rm shear}$, though the half-width-at-half-maximum (HWHM) is $<0.05~{\rm dex}$ in all cases. The kinks evident in the black, blue, and red curves correspond to $T_{\rm max}$ (\tab{sims}).

\begin{figure}
\begin{center}
\includegraphics[trim={0.02cm 0.04cm 0.02cm 0.02cm}, clip, width =0.48 \textwidth]{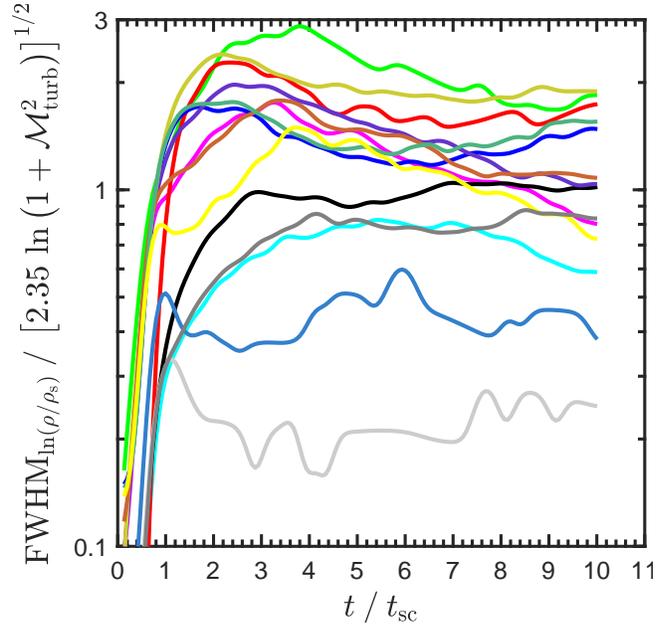}
\end{center}
\caption{FWHM of the density PDFs normalized by the expected value given the turbulent Mach number. Most simulations lie close to unity at all times, implying that the density PDFs are generated by isothermal turbulence. The legend is omitted for clarity, but is identical to previous figures.
}
\label{fig:FWHM_Mach}
\end{figure}

\smallskip
In the bottom-left-hand panel we show the radial profiles of the Hydrogen number density, $n_{\rm H}$. Similar to the temperature profiles, these are qualitatively similar at all times $t>\tsc$. The transition from $n_{\rm H,s}$ to $n_{\rm H,b}$ occurs over a similar region to the temperature transition, with its width similarly increasing with time. In the bottom-right-hand panel, we show the PDFs of density normalized by the initial stream density, $\rhos$, and weighted by stream mass. The distribution of gas with $\rho/\rhos>0.1$ is extremely similar when weighting by total mass, $m$, rather than stream mass, $\psi m$. The time variation of the PDFs at $t>\tsc$ is extremely small. Unlike the narrowness of the temperature PDFs towards low temperatures, the density PDFs are roughly log-normal near their peaks, with overdensities as large as 10 seen in the most rapidly cooling runs. The width of the distribution towards low densities increases with the ratio $\tcm/t_{\rm shear}$. The difference between the shapes of the density and temperature PDFs indicates that pressure equilibrium is not maintained, or that the turbulent pressure is not negligible.

\begin{figure*}
\begin{center}
\includegraphics[trim={0.02cm 0.04cm 0.04cm 0.02cm}, clip, width =0.99 \textwidth]{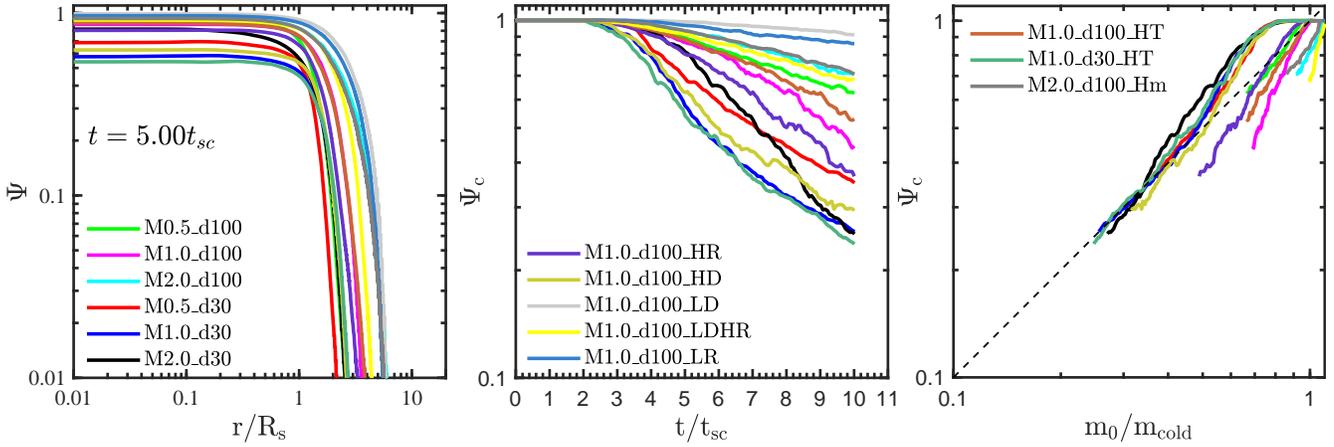}
\end{center}
\caption{Mixing of entrained background material within the stream. \textit{Left:} Radial profiles of the passive scalar $\psi$ at $t=5\tsc$. These are extremely flat within $\Rs$, suggesting efficient mixing within the stream. \textit{Centre:} The central value of the passive scalar, $\psi_{\rm c}=\psi(r=0)$, as a function of time. In all cases where $\tcm<t_{\rm shear}$, $\psi_{\rm c}$ begins declining at $t\sim (2-3)\tsc$, and reaches $\sim (0.25-0.7)$ by $10\tsc$. \textit{Right:} Here we compare $\psi_{\rm c}$ to the ratio of initial stream mass to the total mass of cold gas (see \fig{mass_panels}). The simulations lie close to the one-to-one relation (thin dashed line), suggesting that the entrained background material is efficiently mixed within the stream.
}
\label{fig:col_profile} 
\end{figure*}

\smallskip
A log-normal density PDF is expected for isothermal turbulence \citep[e.g.][]{VS94,Padoan97,Scalo98,Federrath08,Price11,Hopkins12b,Konstandin12}. The width of the distribution is a function of the turbulent Mach number, 
\be 
\label{eq:turb_mach_pdf}
\sigma_{\rm ln(\rho)} \simeq \left[{\rm ln}\left(1+b_{\rm turb}^2\mathcal{M}_{\rm turb}^2\right)\right]^{1/2},
\ee
{\no}where $b_{\rm turb}$ depends on the ratio of compressive to solenoidal forcing driving the turbulence, such that $b_{\rm turb}=1$, $1/3$, or $2/3$ for purely compressive, purely solenoidal, or an equal mixture of compressive and solenoidal forcing, respectively \citep{Federrath08}. While this relation was derived for supersonic turbulence, it is also a good fit for subsonic turbulence, provided the forcing is largely compressive \citep{Konstandin12}. Precisely fitting the density PDFs is beyond the scope of this paper\footnote{In particular, we do not differentiate between a log-normal and a skewed log-normal distribution, see \citet{Vossberg19}.}, as is validating via a power spectrum analysis whether the turbulence in our simulations is fully developed. Nonetheless, we test whether \equ{turb_mach_pdf} is a reasonable description of the simulation data by evaluating the full-width-at-half-maximum (FWHM) of the distributions at each timestep, and using the relation for a log-normal distribution, 
$\sigma = {\rm FWHM}\,[8{\rm ln}(2)]^{-1/2} \simeq {\rm FWHM} / 2.35$. 
When evaluating the FWHM, we use the natural logarithm, ln, rather than the 10-base log shown on the $x$ axis of the bottom-right-hand panel of \fig{temp_dens_panels}, ${\rm ln(\rho)}={\rm log(\rho)}/{\rm log}(e)\simeq 2.30\,{\rm log(\rho)}$. We compare this to \equ{turb_mach_pdf} in \fig{FWHM_Mach}. We assume $b=1$, corresponding to fully compressive forcing, valid if the turbulence is driven primarily by background gas condensing radially onto the stream, rather than by the shear flow. We use the turbulent Mach numbers presented in the left-hand panel of \fig{turb_panels}, i.e. using the initial sound speed of the stream. Although the PDFs appear systematically wider than predicted by \equ{turb_mach_pdf}, for most simulations the discrepancy is less than $\sim 50\%$ at $t>(1-2)\tsc$. This suggests that the density distribution is indeed generated by near-isothermal turbulence with compressive forcing. The simulations which deviate the most from this relation, and which seem to have too narrow a PDF compared to \equ{turb_mach_pdf}, are 
those with $\tcm/t_{\rm shear}\gsim 1$, M1.0\_d100\_LD and M1.0\_d100\_LR. These simulations deviate the most from isothermality as evidenced by the temperature PDFs in \fig{temp_dens_panels}. The larger stream temperature implies that the effective Mach number is lower than assumed in \fig{FWHM_Mach}, which may help bring these curves in better agreement with the prediction.

\subsubsection{Mixing in the Stream}

\Figs{colour_panel_1} and \figss{panels_2} suggest that when $\tcm<t_{\rm shear}$, the turbulence leads to very efficient mixing of the background fluid within the stream. We examine this more quantitatively in \fig{col_profile}. In the left-hand panel we show radial profiles of the passive scalar, $\psi$, at $t=5\tsc$. The profiles are qualitatively similar at all times $t>\tsc$, and are all remarkably constant within $\Rs$, suggesting efficient mixing within the stream. In the centre panel we show $\psi_{\rm c}=\psi(r=0)$ as a function of time. For the LD and LR simulations, where $\tcm \gsim t_{\rm shear}$, this declines by $\sim 10\%$ after $10\tsc$, consistent with the visual impression in \fig{panels_2}. For all other cases, the central value of $\psi$ begins declining at $t\sim (2-3)\tsc$, and is $\sim (0.25-0.70)$ at $t\sim 10\tsc$, suggesting that a significant amount of background material has been deposited near the central axis of the stream. In the right-hand panel we compare $\psi_{\rm c}$ to the ratio of initial stream mass to the current mass of cold gas (\fig{mass_panels}). If the entrained material were perfectly homogeneously mixed, these two quantities would be equal. While not perfectly equal, they do not lie far from the one-to-one relation, implying that turbulence efficiently mixes the entrained background material within the stream.

\subsection{Cooling Emissivity}
\label{sec:emiss}

\begin{figure*}
\begin{center}
\includegraphics[trim={0.02cm 0.09cm 0.02cm 0.01cm}, clip, width =0.99 \textwidth]{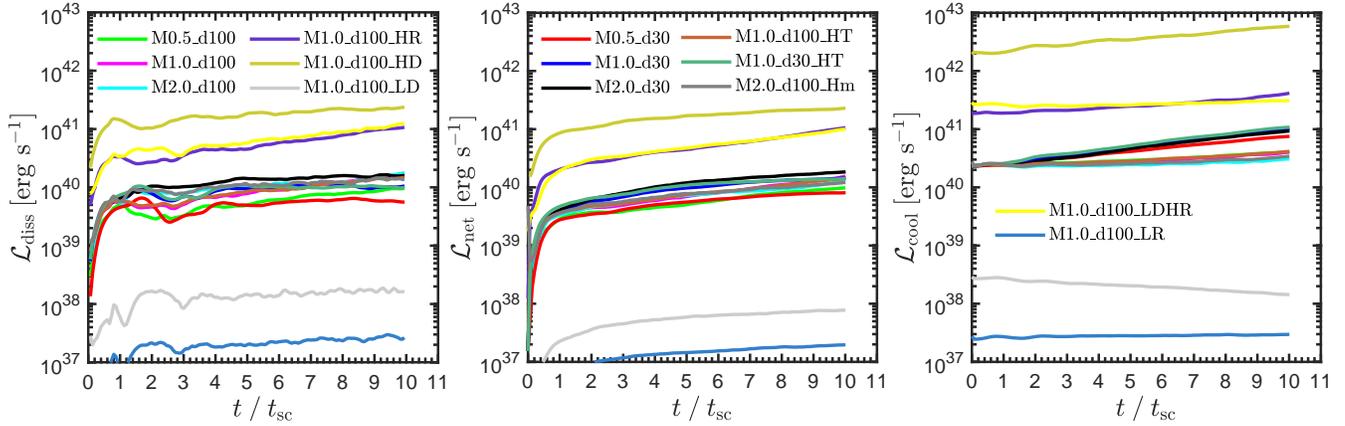}
\end{center}
\caption{Energy dissipation and cooling radiation in the simulations. \textit{Left:} The net energy dissipation rate measured directly from the simulation, accounting for total kinetic energy and thermal enthalpy as well as energy advected across the simulation boundary. \textit{Centre:} Net radiative cooling computed in the simulations at each timestep, after subtracting the radiative heating due to the UV background. This matches the measured energy dissipation remarkably well, and obeys the scaling relations predicted in the text, showing that the dissipation processes described in the previous sections indeed lead to cooling at the predicted rates. \textit{Right:} Total cooling computed in the simulations, without subtracting the radiative heating by the UV background. The cooling induced by the UV background dominates over that induced by the instability, accounting for $\sim (95,\,80,\,50)\%$ of the total radiation for stream densities of $n_{\rm H,s}\sim (0.1,\,0.01,\,10^{-3})\cmc$ respectively. This contribution will likely be reduced when self-shielding of dense gas is included in the models. 
}
\label{fig:lumis_panels} 
\end{figure*}


\smallskip
In \se{decel} and \se{turb}, we quantified the kinetic energy in both laminar flow and turbulent motions, and the thermal energy in both the stream and the background fluids, addressing the transfer of energy between these channels and the loss of energy from the system. We now wish to quantify the cooling emission in our simulations and to compare this to the energy loss associated with the instability.

\smallskip
In the left-hand panel of \fig{lumis_panels}, we show the net energy dissipation rate in our simulations. This is computed by directly measuring the total kinetic (laminar plus turbulent) and thermal (stream plus background) energy in the simulation volume at each output, while also accounting for energy flux through the simulation boundary (negligible compared to other sources of energy loss). We then compute the energy dissipation rate between two adjacent snapshots. When doing this, we multiply the thermal energy loss by $(5/3)$, as expected for an isobaric cooling flow due to the PdV work on the cooling gas compensating for some of the radiated energy \citep[][G20]{Fabian94}. In other words, it is the difference in the enthalpy of the gas between two timesteps which is radiated, rather than the internal energy. This was found to be a good description of the emissivity in G20, and we also find it to accurately describe our simulations, as shown below. This is because the background pressure far from the stream remains roughly constant throughout the simulation, and acts as a pressure bath confining the cooling flow of entrained mass onto the stream. 

\begin{figure}
\begin{center}
\includegraphics[trim={0.02cm 0.04cm 0.02cm 0.02cm}, clip, width =0.48 \textwidth]{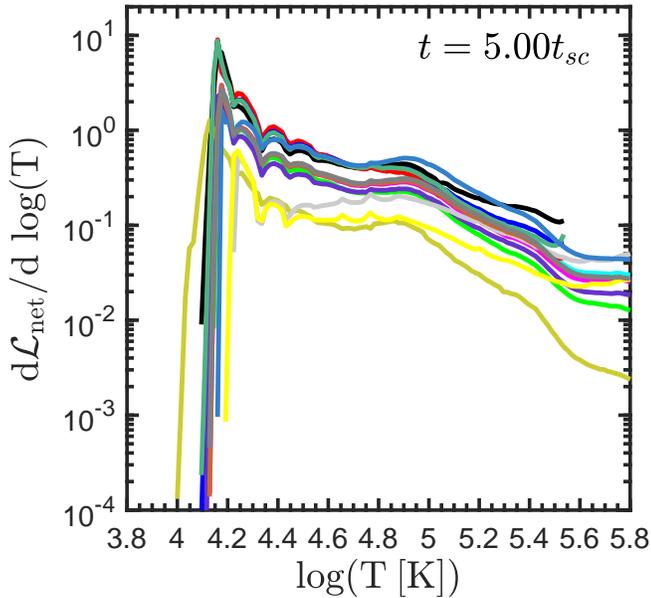}
\end{center}
\caption{PDF of net radiative cooling computed in the simulations, as a function of gas temperature, at $t=5\tsc$. Each curve is normalized to an integral of unity, while the total luminosity can be read from the centre panel of \fig{lumis_panels}. In all cases, the radiation is dominated by gas with $T\sim (2-3)\times 10^4\K$. The legend is omitted for clarity, but is identical to previous figures.
}
\label{fig:lumis_temp_pdf}
\end{figure}

\smallskip
In the centre panel of \fig{lumis_panels} we show the net cooling rate in our simulations, computed directly from the \texttt{RAMSES} cooling modules after subtracting the heating from the UV background. In this way, we focus only on excess cooling beyond what is needed to maintain thermal equilibrium with the UVB. Recall that we shut off cooling for gas with $T>T_{\rm max}$. These agree remarkably well with the dissipation rates shown in the left-hand panel, showing that the energy loss induced by the instability is emitted as cooling radiation. In \fig{lumis_temp_pdf} we show the distribution of this radiation as a function of temperature at $t=5\tsc$, but note that the distributions are extremely similar at all $t>\tsc$. 
In all cases, the emission peaks at $T\sim (1.5-2)\times 10^4 \K$, and $\sim 50\%$ of the emission originates from gas with $T<5\times 10^4 \K$ within the turbulent mixing zone near the stream-background interface. The emission is thus expected to be dominated by Ly$\alpha$ \citep{Goerdt10}. Note that the emission from gas with $T$ greater than the fiducial $T_{\rm max}\sim 10\Ts$ (\tab{sims}) is smaller than gas with $T\lsim 3\times 10^4$ by a factor of $\sim (10-100)$. This explains why increasing $T_{\rm max}$ by a factor of 4 in the two HT runs did not change any of our results (see also G20). 

\smallskip
At late times, the energy loss is dominated by the thermal energy of background gas, given by \equ{erad}. Together with \equ{tent}, this implies that 
\be 
\label{eq:rad_scaling}
\begin{array}{c}
\mathcal{L}_{\rm diss}\propto m_0\, \cb^2\, t_{\rm ent}^{-1}\propto n_{\rm s}\,\Rs^2\, L\, \cb^2\, \delta^{-1}\, t_{\rm cool}^{-1/4}\,\tsc^{-3/4} \\
\\
\propto n_{\rm s}^{5/4}\, \Rs^{5/4}\, L\, \cs^{11/4},
\end{array}
\ee
{\no}where $L$ is the stream length, and we have used the fact that $t_{\rm cool}\propto n^{-1}$, and that $\delta^{-1}\cb^2=\cs^2$. All our simulations have comparable stream temperatures and sound speeds, while the stream length is $L=32\Rs$. We thus obtain $\mathcal{L}_{\rm diss}\propto n_{\rm s}^{5/4} \Rs^{9/4}$. This explains why, in \fig{lumis_panels}, all simulations with $n_{\rm H,s}=0.01\cmc$ and $\Rs=3\kpc$ have roughly the same luminosity, while the luminosity in the HR simulation with $\Rs=6\kpc$ (purple line) is a factor of $\sim 5$ larger, and that in the HD simulation with $n_{\rm H,s}=0.1\cmc$ (yellow line) is a factor of $\sim 20$ larger. The luminosity in the LD, LR, and LDHR simulations are lower than expected from this scaling. These have the largest ratios of $\tcm/t_{\rm shear}$ among our simulations (\tab{sims}), so the entrainment of background gas onto the stream is suppressed and the scaling relations derived above do not strictly apply. In a companion paper \citep{M20}, we use these results to estimate the total luminosity emitted by a cosmological cold stream in a dark matter halo potential as a result of these dissipation processes.

\smallskip
In the right-hand panel of \fig{lumis_panels}, we show the total cooling luminosity in the simulations, computed from the \texttt{RAMSES} cooling modules without subtracting the radiative heating by the UVB. Comparing this to the centre panel, it is clear that without self-shielding the cooling induced by the UVB dominates over the dissipation processes described above. The latter comprise $\sim 50\%$, $20\%$, and $5\%$ of the total emission for $n_{\rm H,s}\sim 10^{-3}$, $10^{-2}$, and $10^{-1}\cmc$, respectively. The increased emission due to UV heating is primarily from gas near the thermal equilibrium temperature with $T\sim \Ts$ in the stream interior. The large overdensities in the cold gas (bottom-right-hand panel of \fig{temp_dens_panels}) yield particularly large emissivity in this regime, especially in the HD simulation. Metal line cooling at $T<10^4$ has a negligible contribution to the total emissivity. We defer a more detailed exploration of the resulting spectrum, and the relative importance of UV heating versus KHI induced dissipation, to future work which will include self-shielding (see \se{phys}).

\section{Caveats and Additional Physics}
\label{sec:phys} 

\smallskip
While our analysis has been thorough in terms of the effect of radiative cooling on stream evolution, there are many physical processes relevant for astrophysical cold streams which are still unaccounted for in our models. Some of these have been addressed by other authors, though not in combination with radiative cooling. In this section we address these additional processes, and speculate how they may influence our results. 

\smallskip
Before addressing additional physics, we note that \citet{Vossberg19} used 2d purely hydrodynamic simulations to study the amplitude of KHI-induced density fluctuations within cold streams. They found the resulting density fluctuations both short lived and small compared to those inferred from observations of the CGM of galaxies expected to host cold streams, and concluded that additional physical processes, such as cooling or gravity, were necessary to achieve such large density fluctuations. As we have shown, radiative cooling enhances the density fluctuations within streams, and prolongs their lifetime since the stream density is not diluted by the KHI (\figs{density_panel_1}, \figss{panels_2}, and \figss{temp_dens_panels}). However, these fluctuations are still small compared to those inferred from observations. In some cases, the lognormal dispersion is inferred to be as large as $\sigma_{\rm ln(\rho)}\sim 2.5$ \citep{Cantalupo19}. This implies 
${\rm FWHM}_{\rm log(\rho)}\sim 2.6$, much larger than seen in our simulations (\fig{temp_dens_panels}, bottom-right). It thus seems that if KHI in cold streams is to explain the large density fluctuations observed, additional physical processes beyond radiative cooling are required.

\smallskip
One such process may be self-shielding. 
The cold streams are expected to be largely self-shielded from the UV background, as discussed in \citet{FG10} and \citet{Goerdt10}, who studied the emission properties of cold streams using cosmological simulations. These two studies disagreed on the density above which self-shielding becomes important, with estimates ranging from $n_{\rm H}\sim (0.01-0.1)\cmc$, causing their estimates for the Ly$\alpha$ emissivity in cold streams to vary by more than an order of magnitude. As we have shown, in the limit of no self-shielding the cooling radiation induced by the UV background can be much larger than that induced by the instability (\fig{lumis_panels}). 
Furthermore, as self-shielding affects the thermal and ionization state of the gas, it must be accounted for in order to properly estimate the fraction of radiation emitted in Ly$\alpha$, even only considering radiation induced by the instability. Finally, self-shielding is likely to widen the density distribution within the streams. The overdensities of a factor $\sim (3-10)$ produced in our current simulations may be able to cool below $10^4\K$, especially given the efficient mixing of background metals in the stream (\fig{col_profile}), thereby reaching even higher densities. We will include self-shielding in our models in an upcoming paper.

\smallskip
Another important process affecting the density distribution and morphology of streams is self-gravity. The importance of self-gravity in streams can be assessed by comparing the mass per unit length of the stream, $\lambda=\pi \Rs^2 \rhos$, to the maximal mass per unit length possible for an isothermal stream in hydrostatic equilibrium \citep{Ostriker64},
\be 
\label{eq:lambda_max}
\lambda_{\rm max}=2\cs^2/G \simeq 10^5\msun\,\pc^{-1}~T_4\,\mu_{\rm s,0.6}^{-1},
\ee 
{\no}where $G$ is Newton's gravitational constant, and $\mu_{\rm s,0.6}=\mu_{\rm s}/0.6$, with $\mu_{\rm s}$ the mean molecular weight in the stream. If the stream is strongly self-shielded and predominantly neutral, $\mu_{\rm s,0.6}\sim 2$. The mass per unit length of the stream can be constrained via the mass accretion rate through the stream (P18; \citealp{M18,M20}), ${\dot{M}}_{\rm s}\sim \pi \Rs^2 \rhos V_{\rm s}\sim f_{\rm s}f_{\rm b}{\dot{M}}_{\rm v}$, where $V_{\rm s}= \eta \Vv$ is the stream velocity which is of order the virial velocity (i.e. $\eta\sim 1$), ${\dot{M}}_{\rm v}$ is the total mass accretion rate through the virial radius, $f_{\rm b}$ is the Universal baryon fraction, and $f_{\rm s}$ is the fraction of the total accretion along a given stream. Note that we have assumed here that the baryonic accretion is predominantly gas, which is a reasonable assumption at high redshift. Cosmological simulations suggest $f_{\rm s}\sim (0.2-0.5)$ with a typical value of $f_{\rm s}=1/3$ \citep{Danovich12}. In the Einstein de Sitter (EdS) regime (valid at $z>1$), the accretion onto the virial radius is well approximated by \citep{Dekel09,Dekel13} ${\dot{M}}_{\rm v}\sim 468\sy\,M_{12}\,(1+z)_3^{5/2}$, with $M_{12}=\Mv/10^{12}\msun$ and $(1+z)_3=(1+z)/3$. The virial velocity is \citep{Dekel13} $\Vv \sim 200\kms\,M_{12}^{1/3}\,(1+z)_3^{1/2}$. Taking all this together, we have
\be 
\label{eq:grav_model}
\lambda/\lambda_{\rm max}\simeq 1.2~M_{12}^{2/3}\,(1+z)_3^2\,\widetilde{s}\,\eta^{-1}\,T_4^{-1}\,\mu_{\rm s,0.6},
\ee 
{\no}where $\widetilde{s}\sim (0.3-3)$ accounts for halo-to-halo variance in the normalization of ${\dot{M}}_{\rm v}$ and $f_{\rm s}$. This implies that for a wide range of stream parameters hydrostatic equilibrium may not be possible. This was discussed by \citet{M18}, who speculated that this could lead to gravitational collapse and star formation in streams within the CGM, and potentially even to the formation of globular clusters at high redshift. Heating by a UV background yields $T_4\gsim (1.5-2)$ (\tab{sims}) which can help stabilize the stream. However, self-shielding will enhance the instability, by allowing the stream to cool below $10^4\K$, reducing $T_{\rm 4}$ and increasing $\mu_{\rm s,0.6}$ \citep[see also][]{Li_hopkins19}.

\smallskip
If $\lambda<\lambda_{\rm max}$, self-gravity can still influence the development of KHI in the streams. This regime was studied by \citet{Aung19}, who found that if $\lambda$ was smaller than a critical value $\lambda_{\rm crit}<\lambda_{\rm max}$ (which is larger for denser and faster streams), then stream deceleration proceeds very similarly to the non-gravitating case studied in M19, but total stream disruption is slowed by a factor of $\gsim 3$ due to buoyancy forces near the stream core, even for very low values of $\lambda$. This allows streams $\gsim 3$ times thinner than the M19 estimate to reach the central galaxy, irrespective of radiative cooling. For $\lambda_{\rm crit}<\lambda<\lambda_{\rm max}$, the stream was found to fragment into long-lived, dense, pressure-confined clumps. Furthermore, self-gravity will likely enhance the large overdensities we find in our simulations with cooling, and these may trigger local gravitational instabilities, in addition to the global fragmentation discussed in \citet{M18} and \citet{Aung19}. We will study the combined effects of cooling and gravity in an upcoming paper (Cornuault et al, in prep.).

\smallskip
Thermal conduction can hinder both hydrodynamic and thermal instabilities, and may be important for the evolution of a multiphase medium \citep[e.g.][]{Armillotta16,Armillotta17,Li_hopkins19}. In particular, this may influence whether the hot gas cools and condenses onto the stream, as in our simulations, or whether the cold stream diffuses and mixes into the background. For a cold cloud embedded in a hot medium, radiative processes dominate over thermal conduction if the cloud is larger than the Field Length \citep{Begelman_Mckee90,Armillotta16}, 
\be 
\label{eq:LField}
L_{\rm Field}=\left(\frac{\kappa_{\rm Sp}T_{\rm h}}{n_{\rm c}^2\Lambda(T_{\rm c})}\right)^{1/2},
\ee
{\no}where 
\be 
\label{eq:Spitzer}
\kappa_{\rm Sp}\simeq \frac{1.84\times 10^{-5}}{{\rm ln}(\Psi)}T_{\rm h}^{5/2}~{\rm erg\,s^{-1}\,\K^{-1}\,cm^{-1}} 
\ee 
{\no}is the Spitzer heat conduction coefficient and
\be 
\label{eq:coulomb}
{\rm ln}(\Psi)\simeq 29.7+{\rm ln}\left[\frac{T_{\rm e}/10^6\K}{\sqrt{n_{\rm e}/\cmc}}\right] 
\ee 
{\no}is the Coulomb logarithm, with $T_{\rm e}$ and $n_{\rm e}$ the electron temperature and density in the hot phase. For cold streams, using $T_{\rm e}\sim 10^6\K$ and $n_{\rm e}\sim 5\times 10^{-5}\cmc$, which are typical conditions for the hot CGM gas near $\Rv$ in a $\sim 10^{12}\msun$ halo at $z\sim 2$, we obtain ${\rm ln}(\Psi)\sim 35$. Inserting this and \equ{Spitzer} into \equ{LField}, along with $T_{\rm h}=10^6\K\,\delta_{100}\,T_4$ with $\delta_{100}=\delta/100$ and $T_4=\Ts/10^4\K$, yields 
\be 
\label{eq:LField2}
L_{\rm Field} \simeq 0.2\kpc\,\frac{\delta_{100}^{7/4}\,T_4^{7/4}}{\Lambda_{-23}^{1/2}\,n_{s,0.01}},
\ee 
{\no}where $\Lambda_{-23}=\Lambda(\Ts)/10^{-23}{\rm erg~s^{-1}~cm^3}$, normalized to the approximate cooling rate at $T\sim 1.5\Ts$ in the presence of a $z=2$ UV background as we have assumed throughout. $L_{\rm Field}$ is thus comparable to $R_{\rm s,crit}$ (\equnp{Rscrit}), below which cooling is unimportant irrespective of thermal conduction. Even if $\Lambda_{-23}\sim 0.1$, which may be the case if the streams are self-shielded and $\Ts \lsim 10^4\K$, $L_{\rm Field}$ is still small compared to the radius of a typical stream (\equnp{Rs_Rv}). We conclude that thermal conduction is unlikely to affect our main conclusions regarding stream evolution in the presence of cooling, though it may influence the late time shattering of the stream into dense cloudlets seen in \fig{density_panel_1}.

\smallskip 
When considering thermal conduction, we must also consider magnetic fields, which can significantly lower the thermal conductivity below the Spitzer value, particularly perpendicular to the shear direction, and reduce the Field length even further. On the other hand, magnetic fields may stabilize the stream against KHI and prevent mixing of the stream and background gas irrespective of thermal conduction. \citet{Berlok19b} studied the influence of magnetic fields on the development of KHI in cold streams, while also carefully accounting for the influence of a smooth transition between the stream and the background following \citet{Berlok19a}. They find that while magnetic fields can enhance the linear growth rates of the instability, they inhibit stream disruption by suppressing the mixing of the stream and background gas. They estimate that magnetic field strengts of $(0.3-0.8)\mu {\rm G}$ allow streams to survive for $\sim (2-8)$ times longer than the non-magnetic case. It is interesting that radiative cooling, self-gravity, and magnetic fields all seem to prolong the lifetime of streams when acting alone, though for different reasons. For example, magnetic fields prevent the mixing of hot and cold gas thus keeping the stream dense and collimated, while it is precisely this mixing which allows radiative cooling to maintain high densities long-lived streams. In future work, we will study the effects of these physical processes acting in unison on the lifetime and morphology of cold streams.

\smallskip
The gravitational potential of the dark matter halo into which the stream are flowing will have three main effects: (1) the streams will be accelerated towards the halo centre which can counteract the KHI induced deceleration, (2) there will be a gradient in density and pressure in both the background gas and the stream gas, and (3) the stream cross section will decrease with halocentric radius due to both the external pressure gradient as well as pure gravitational focusing. In a companion paper \citep{M20} we present a toy model attempting to account for these processes analytically, and to estimate the resulting profile of stream velocity, mass entrainment, and emitted luminosity. In the future, we also plan to study this directly using idealized simulations of streams embedded in an external potential. 

\smallskip
Finally, several groups have recently performed cosmological simulations with enhanced refinement in the CGM \citep{Hummels18,Peeples19,vdv19}. While the galaxies simulated in these works are less massive than those considered here and may not host a hot static halo at $z\gsim 2$, and the resolution in the CGM is still lower than in our idealized simulations by a factor of a few, we may soon be able to study stream evolution with adequate spatial resolution in fully cosmological simulations. This will allow for self-consistent generation of linear and non-linear perturbations, the interactions of streams with outflows from the central galaxy and with satellite galaxies along the streams, and the complex interaction between multiple streams and the central galaxy in the inner halo, all of which are likely important for the eventual fate of cold streams, and the emission they may produce.

\section{Summary and Conclusions}
\label{sec:conc} 

\smallskip
Massive halos of $\Mv\gsim 10^{12}\msun$ at redshift $z\gsim 2$ are thought to be fed by cold streams with temperatures $\Ts\gsim 10^4\K$. These streams flow along cosmic web filaments and penetrate the hot CGM of these halos, with $T_{\rm h}\sim 10^6\K$. The streams have Mach numbers of $\Mb\sim (0.75-2.25)$ with respect to the halo sound speed, density contrasts of $\delta\sim (30-300)$ compared to the background density (assuming pressure equilibrium with the halo gas), radii $\Rs \sim (0.03-0.30)$ times the halo virial radii, and Hydrogen number densities in the range $n_{\rm H,s}\sim (0.1-5)\times 10^{-2}\cmc$. In order to study the evolution of such streams as they interact with the hot CGM on their way towards the central galaxies, we presented here a detailed analysis of the non-linear stages of Kelvin-Helmholtz Instability (KHI) in the presence of radiative cooling and heating by a UV background, though neglecting for the time being self-shielding and thermal conduction. This extended our previous work on the purely hydrodynamic case \citep{M16,M19,P18}, and on the self-gravitating case \citep{Aung19}, though our current models do not account for gravity. Our main results can be summarised as follows:

\begin{enumerate}

    \item The key parameter in determining the fate of the streams is the ratio $\tcm/t_{\rm shear}$, where $\tcm$ is the cooling time in the turbulent mixing layer which forms at the stream-background interface as a result of the instability, and $t_{\rm shear}$ is the timescale for the shear layer to grow to the width of the stream in the non-radiative case analysed by \citet{M19}. If $\tcm>t_{\rm shear}$, the behaviour is very similar to the non-radiative case, the stream expands and mixes into the background, its density is diluted, and it is eventually disrupted (\fig{panels_2}). If $\tcm<t_{\rm shear}$, on the other hand, KHI does not disrupt the stream. Rather, background gas cools and condenses onto the stream, increasing the cold gas mass by up to a factor of $\sim 3$ within 10 stream sound crossing times, though for $0.1 t_{\rm shear}\lsim \tcm\lsim t_{\rm shear}$, the entrainment rates are a factor of a few lower (\fig{mass_panels}). In this regime, the stream remains cold, dense and collimated until very late times, when it may fragment into small cloudlets (\figs{density_panel_1}-\figss{panels_2}), either due to thermal instabilities \citep{McCourt18} or secondary instabilities following the development of long wavelength perturbations \citep{M19}. Similar behaviour was found for cold clouds accelerating in a hot wind \citep{Gronke18,Gronke20,Li_hopkins19}.
    
    \item The condition $\tcm = t_{\rm shear}$ can be translated to a critical stream radius, $\Rs = R_{\rm s,crit}$, such that $\Rs/R_{\rm s,crit} = t_{\rm shear}/\tcm$ (\fig{Rs_crit}). For realistic cold streams feeding hot halos, $\Rs>R_{\rm s,crit}$ except for very extreme parameters, and nearly irrespective of the metalicity in the stream (\equsnp{Rscrit}-\equmnp{Rvir}). This implies that it is very difficult for such streams to be disrupted by hydrodynamic instabilities in the CGM of massive halos before reaching the central galaxy. Moreover, $R_{\rm s,crit}$ is comparable to the Field length assuming Spitzer conductivity (\equnp{LField2}), so thermal conduction is unlikely to affect these conclusions.
    
    \item The entrained background gas rapidly mixes with the stream gas (\figs{colour_panel_1}, \figss{panels_2}, \figss{col_profile}). This causes the stream to decelerate and loose bulk kinetic energy as a result of momentum conservation (\fig{decel_panels}). However, for $\Mb\lsim 1$, the kinetic energy loss is small, $\lsim 30\%$, compared to the thermal energy lost by the background gas as it condenses onto the stream (\fig{EK_over_Eth}). For $\Mb\sim 2$, or for cases where $\tcm>t_{\rm shear}$, the bulk kinetic energy losses become comparable to or can exceed the thermal energy losses.
    
    \item In steady state, roughly $\sim (1-10)\%$ of the bulk kinetic energy losses up to that time are maintained as excess thermal energy within the stream (\fig{Eths}). The remainder is radiated away, together with the thermal enthalpy losses of the background gas. The turbulent velocities in the mixing layer are $\sigma_{\rm turb}\sim 0.2 V_{\rm s,0}\delta^{-1/2}$, corresponding to turbulent Mach numbers of $\mathcal{M}_{\rm turb}\sim (0.1-0.5)$ with respect to the cold component (\fig{turb_panels}), or velocities of $\sigma_{\rm turb}\sim (2-10)\kms$. Such a stream detected in absorption would be dominated by thermal broadening, consistent with a claimed detection of a cold stream in absorption around a massive star-forming galaxy at $z\sim 2.5$ \citep{Crighton13}.
    
    \item The turbulence appears to be driven by primarily compressive forcing, resulting from the radial condensation of background gas onto the stream. This induces a roughly lognormal density distribution in the stream, whose width follows the predicted relation with turbulent Mach number (\figs{temp_dens_panels} - \figss{FWHM_Mach}; \citealp{Konstandin12}). Overdensities of a factor $\sim (3-10)$ are not uncommon in the streams, even at late times, though the distribution is still narrow compared to that inferred from observations \citep{Cantalupo19}. The temperature distribution, on the other hand, remains sharply peaked around the initial stream temperature, with very little gas at lower temperatures. This is a result of heating from the UV background, and is likely to change once self-shielding of dense gas is accounted for. This is also likely to broaden the density PDF, particularly in conjunction with self-gravity \citep{Aung19}, possibly bringing the density fluctuations more in line with observations. 
    
    \item The total cooling emission from our model streams is dominated by UV background heating. This comprises $\sim (50-95\%)$ of the total emission, depending on stream density. However, this will almost certainly change when self-shielding is included in our models. The excess cooling radiation, beyond that induced by the UV background, is in excellent agreement with the net energy dissipation rates predicted by our model (\fig{lumis_panels}). Roughly half of the emissivity originates from gas with $T< 5\times 10^4\K$ within the turbulent mixing layer near the stream interface (\fig{lumis_temp_pdf}), despite the fact that most of the energy loss is the thermal energy of the hot background, with $T\sim 10^6\K$. The emission is thus expected to be dominated by Ly$\alpha$. We present predictions for the total emitted luminosity of cold streams in dark matter halos in a companion paper \citep{M20}, though detailed predictions of the emission spectrum await the inclusion of self-shielding in our models. 
    
\end{enumerate}

\section*{Acknowledgments} 
We thank the anonymous referee for a constructive and substantive report which helped improve the quality and accuracy of this manuscript. We thank Romain Teyssier for making \texttt{RAMSES} 
publicly available, and for his many helpful suggestions when running the simulations. We thank Thomas Berlok, Andi Burkert, Nicolas~Cornuault, Drummond Fielding, Max Gronke, Joseph F.~Hennawi, Suoqing~Ji, Neal Katz, S. Peng Oh, Christoph~Pfrommer,~X.~Prochaska, Santi Roca-Fabrega, and~Chuck~Steidel for helpful discussions. The simulations were performed on the Grace cluster at Yale and the magny cluster at HITS. NM and FCvdB acknowledge support from the Klauss Tschira Foundation through the HITS Yale Program in Astropysics (HYPA). AD is partly supported by the grants NSF AST-1405962, BSF 2014-273, GIF I-1341-303.7/2016, and DIP STE1869/2-1 GE625/17-1. YB acknowledges ISF grant 1059/14. FCvdB received additional support from the National Aeronautics and Space Administration through grant No. 17-ATP17-0028 issued as part of the Astrophysics Theory Program. 

\bibliographystyle{mn2e} 

\appendix

\section{Convergence Tests}
\label{sec:convergence}

\begin{figure}
\begin{center}
\includegraphics[trim={0.02cm 0.05cm 0.02cm 0.01cm}, clip, width =0.445 \textwidth]{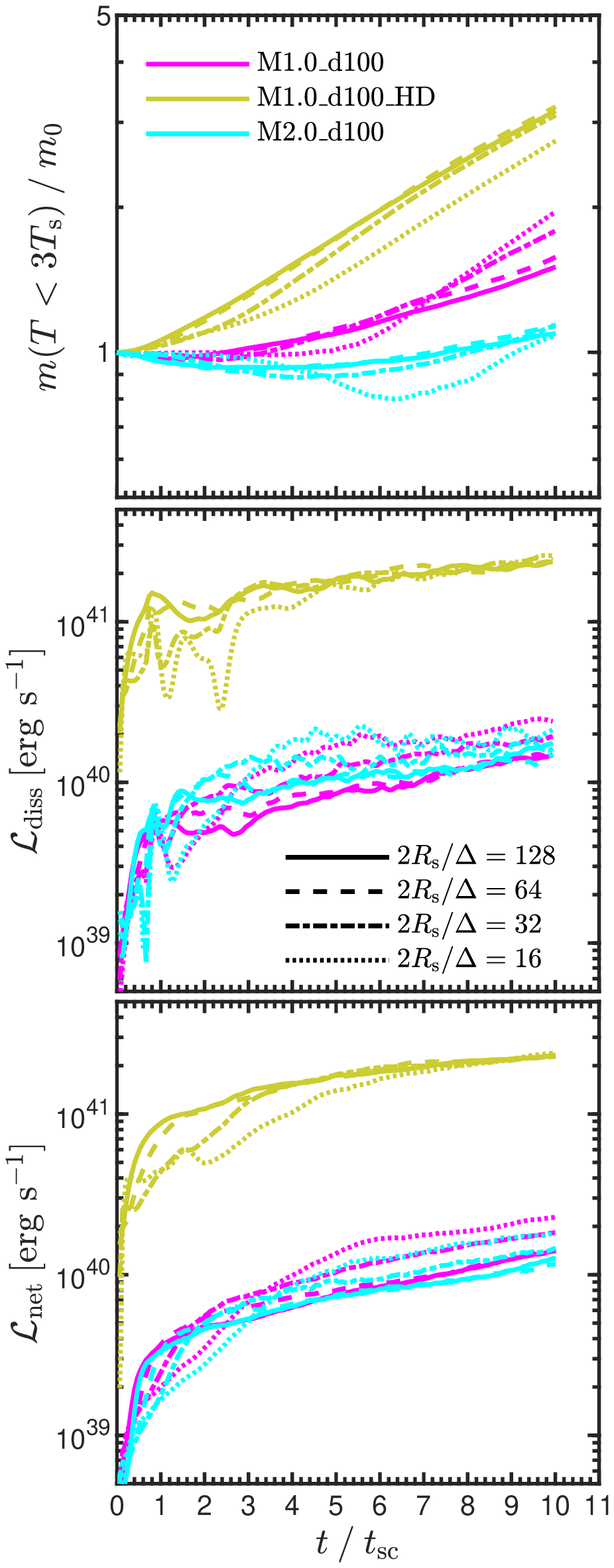}
\end{center}
\caption{Convergence tests. We show as a function of time normalized by the stream sound crossing time, the mass of cold gas normalized by the initial stream mass (\textit{left}, similar to \fig{mass_panels} left), the net energy dissipation (\textit{centre}, similar to \fig{lumis_panels} left), and the net cooling rate (\textit{right}, similar to \fig{lumis_panels} centre). Different colours represent different simulations as in \figs{mass_panels}-\figss{lumis_panels}, while different line styles represent different resolutions, as indicated in the legend. Our results appear converged even at very low resolution of 16 cells across the stream diameter.}
\label{fig:converge} 
\end{figure}

\smallskip
We repeated three of our simulations, M1.0\_d100, M2.0\_d100, and M1.0\_d100\_HD, with several different resolutions, ranging from 128 cells across the stream diameter (our fiducial resolution) to 16 cells across the stream diameter. We present the results of this study in \fig{converge}. In the left-hand panel we show the cold mass as a function of time, similar to the left-hand panel of \fig{mass_panels}. In the centre and right-hand panels we show the the net energy dissipation rate and the net cooling rate in our simulations, similar to the left-hand and centre panels of \fig{lumis_panels}. Note that since the energy dissipation is governed by the mass entrainment, convergence in the left-hand panel implies convergence in the centre and right-hand panels, but we show all three for completeness. 
Our results are approximately converged even at very low resolutions, with only 16 cells across the stream diameter. For the  M1.0\_d100 and M2.0\_d100 cases, There is a slight tendency for the entrainment rate and the energy dissipation rate to be higher in lower resolution simulations, but this is a small effect. G20 reported similar convergence properties of the mass entrainment rate in their cloud crushing simulations with radiative cooling. 

\label{lastpage} 
 
\end{document}